\documentstyle[12pt]{article}

\catcode`@=11
\def\un#1{\relax\ifmmode\@@underline#1\else
        $\@@underline{\hbox{#1}}$\relax\fi}
\catcode`@=12

\def\a{\alpha}
\def\b{\beta}
\def\c{\chi}
\def\d{\delta}
\def\e{\epsilon}
\def\f{\phi}
\def\g{\gamma}
\def\h{\eta}

\def\j{\psi}

\def\m{\mu}
\def\n{\nu}

\def\p{\pi}
\def\q{\theta}
\def\r{\rho}
\def\s{\sigma}

\def\x{\xi}

\def\F{\Phi}
\def\G{\Gamma}
\def\J{\Psi}
\def\L{\Lambda}

\def\P{\Pi}

\def\S{\Sigma}
\def\U{\Upsilon}
\def\X{\Xi}

\def\ve{\varepsilon}
\def\vf{\varphi}

\def\bo{{\raise-.5ex\hbox{\large$\Box$}}}               
\def\pa{\partial}                                       
\def\pr{\prod}                                          
\def\TH{{\raise.2ex\hbox{$\displaystyle \bigodot$}\mskip-4.7mu \llap H \;}}
\def\face{{\raise.2ex\hbox{$\displaystyle \bigodot$}\mskip-2.2mu \llap {$\ddot
        \smile$}}}                                      

\def\sp#1{{}^{#1}}                              
\def\slash#1{\rlap{\hbox{$\mskip 1 mu /$}}#1}      
\def\Bar#1{\overline{#1}}                       
\def\ket#1{\left| #1\right\rangle}              
\def\VEV#1{\left\langle #1\right\rangle}        
\def\leftrightarrowfill{$\mathsurround=0pt \mathord\leftarrow \mkern-6mu
        \cleaders\hbox{$\mkern-2mu \mathord- \mkern-2mu$}\hfill
        \mkern-6mu \mathord\rightarrow$}
\def\dvec#1{\vbox{\ialign{##\crcr
        \leftrightarrowfill\crcr\noalign{\kern-1pt\nointerlineskip}
        $\hfil\displaystyle{#1}\hfil$\crcr}}}           

\def\frac#1#2{{\textstyle{#1\over\vphantom2\smash{\raise.20ex
        \hbox{$\scriptstyle{#2}$}}}}}                   
\def\ha{\frac12}                                        
\def\sfrac#1#2{{\vphantom1\smash{\lower.5ex\hbox{\small$#1$}}\over
        \vphantom1\smash{\raise.4ex\hbox{\small$#2$}}}} 
\def\bfrac#1#2{{\vphantom1\smash{\lower.5ex\hbox{$#1$}}\over
        \vphantom1\smash{\raise.3ex\hbox{$#2$}}}}       
\def\afrac#1#2{{\vphantom1\smash{\lower.5ex\hbox{$#1$}}\over#2}}    

\def\[{\lfloor{\hskip 0.35pt}\!\!\!\lceil}
\def\]{\rfloor{\hskip 0.35pt}\!\!\!\rceil}

\def\fracm#1#2{\hbox{\large{${\frac{{#1}}{{#2}}}$}}}
\def\half{{\fracm12}}
\def\ha{\half}

\def\un{\underline}
\def\fracmm#1#2{{{#1}\over{#2}}}

\def\low#1{{\raise -3pt\hbox{${\hskip 0.75pt}\!_{#1}$}}}

\newskip\humongous \humongous=0pt plus 1000pt minus 1000pt
\def\caja{\mathsurround=0pt}
\def\eqalign#1{\,\vcenter{\openup2\jot \caja
        \ialign{\strut \hfil$\displaystyle{##}$&$
        \displaystyle{{}##}$\hfil\crcr#1\crcr}}\,}
\newif\ifdtup

\def\ref#1{$\sp{#1)}$}

\def\pl#1#2#3{Phys.~Lett.~{\bf {#1}B} (19{#2}) #3}
\def\np#1#2#3{Nucl.~Phys.~{\bf B{#1}} (19{#2}) #3}
\def\prl#1#2#3{Phys.~Rev.~Lett.~{\bf #1} (19{#2}) #3}
\def\pr#1#2#3{Phys.~Rev.~{\bf D{#1}} (19{#2}) #3}
\def\cqg#1#2#3{Class.~and Quantum Grav.~{\bf {#1}} (19{#2}) #3}

\def\mpl#1#2#3{Mod.~Phys.~Lett.~{\bf A{#1}} (19{#2}) #3}

\parskip=\medskipamount                 
\thicklines                         

\begin{document}


\def\ua{\uparrow}
\def\da{\downarrow}
\def\sdot{\!\cdot\!}
\def\pc{picture-changing}
\def\pco{picture-changing operator}


\thispagestyle{empty}               

\def\border{                                            
        \setlength{\unitlength}{1mm}
        \newcount\xco
        \newcount\yco
        \xco=-24
        \yco=12
        \begin{picture}(140,0)
        \put(-20,11){\tiny Institut f\"ur Theoretische Physik Universit\"at
Hannover~~ Institut f\"ur Theoretische Physik Universit\"at Hannover~~
Institut f\"ur Theoretische Physik Hannover}
        \put(-20,-241.5){\tiny Institut f\"ur Theoretische Physik Universit\"at
Hannover~~ Institut f\"ur Theoretische Physik Universit\"at Hannover~~
Institut f\"ur Theoretische Physik Hannover}
        \end{picture}
        \par\vskip-8mm}

\def\headpic{                                           
        \indent
        \setlength{\unitlength}{.8mm}
        \thinlines
        \par
        \begin{picture}(29,16)
        \put(75,16){\line(1,0){4}}
        \put(80,16){\line(1,0){4}}
        \put(85,16){\line(1,0){4}}
        \put(92,16){\line(1,0){4}}

        \put(85,0){\line(1,0){4}}
        \put(89,8){\line(1,0){3}}
        \put(92,0){\line(1,0){4}}

        \put(85,0){\line(0,1){16}}
        \put(96,0){\line(0,1){16}}
        \put(79,0){\line(0,1){16}}
        \put(80,0){\line(0,1){16}}
        \put(89,0){\line(0,1){16}}
        \put(92,0){\line(0,1){16}}
        \put(79,16){\oval(8,32)[bl]}
        \put(80,16){\oval(8,32)[br]}

        \end{picture}
        \par\vskip-6.5mm
        \thicklines}

\border\headpic {\hbox to\hsize{
\vbox{\noindent DESY 94--104 \\
ITP--UH--05/94 \hfill June 1994 \\
hep-th/9406101 \hfill revised September 1994}}}

\noindent
\vskip1.3cm
\begin{center}

{\Large\bf The GSO Projection, BRST Cohomology and
\vglue.1in
           Picture-Changing in $~N{=}2~$ String Theory}
\footnote{Supported in part by the `Deutsche Forschungsgemeinschaft'}\\
\vglue.3in

Jan Bischoff,
Sergei V. Ketov \footnote{
On leave of absence from:
High Current Electronics Institute of the Russian Academy of Sciences,
Siberian Branch, Akademichesky~4, Tomsk 634055, Russia}
and Olaf Lechtenfeld

{\it Institut f\"ur Theoretische Physik, Universit\"at Hannover}\\
{\it Appelstra\ss{}e 2, 30167 Hannover, Germany}\\
{\sl bischoff, ketov, lechtenf @itp.uni-hannover.de}
\end{center}
\vglue.2in
\begin{center}
{\Large\bf Abstract}
\end{center}

We investigate in detail the critical $N{=}2$ fermionic string with and
without a global ${\bf Z}_2$ twist. An analysis of BRST cohomology shows
that twisted sectors contain massless `spacetime' fermions which are
{\it non-local\/} with respect to the standard massless boson.
However, two distinct GSO projections exist, one (untwisted) retaining
merely the usual boson and its spectral-flow partner, the other (twisted)
yielding two fermions and one boson, on the massless level.
The corresponding chiral BRST-invariant vertex operators are constructed
in certain pictures, and their fusion and picture-changing are investigated,
including the construction of inverse picture-changing operators.
The $N{=}2$ `spacetime supersymmetry' generators are {\it null\/} operators,
since the twisted massless states fail to interact.  The untwisted three-
and four-point functions are recalculated at tree-level.

\newpage
\hfuzz=10pt

\section{Introduction}

The {\it critical} $N{=}2$ fermionic string \cite{aba,aba2} has a gauged
$N{=}2$ superconformal symmetry on its world sheet, it naturally lives in
$4{+}0$ or $2{+}2$ dimensions \cite{ft,ali}, it usually has only a finite
number of physical states \cite{bien}, and it is known to be closely related
with the self-dual four-dimensional field theories and integrable
models~\cite{ov}.~\footnote{See refs.~\cite{ma,ket} for a review.}
The $N{=}0$ (bosonic) and $N{=}1$ (fermionic) strings can also be considered
in the framework of the general $N{=}2$ string theory as particular vacua
\cite{bv,ff,op,bop}.

As was indirectly argued in ref.~\cite{siegel}, the $N{=}2$ string theory
may also support an extended supersymmetry in the target space.
Therefore, this string theory is expected to be related with the extended {\it
supersymmetric} self-dual gauge theories and extended self-dual supergravities
in $2{+}2$ dimensions \cite{kng}. The main objection against a presence of
`spacetime' supersymmetry seemed to be the apparent absence of `spacetime'
{\it fermionic} states in the physical spectrum of the (untwisted) $N{=}2$
string. Recently, two of us \cite{klp} proposed a possible resolution to
this problem, by considering non-trivial background topologies which add
{\it twisted} sectors to the $N{=}2$ string.
We have shown that there is room for physical `spacetime'
fermions in the $N{=}2$ string theory by explicitly constructing the
BRST-invariant (physical) chiral vertex operators corresponding to the bosonic
{\it and} fermionic physical states which appear as the massless ground states
in certain ${\bf Z}_2$-twisted sectors of the theory. The number of `spacetime'
fermions, if any, depends on the global topology of `spacetime'. In
the case of the flat half-space  ${\bf C}^{1,1}/{\bf Z}_2$, with ${\bf Z}_2$
representing complex conjugation, we were lead to just two fermionic states.
In ref.~\cite{klp} we have confined ourselves to a {\it
free} $N{=}2$ string, since non-local OPEs between bosonic and fermionic chiral
vertex operators prevented us to define interactions between physical bosons
and fermions. We also suggested that an {\it asymmetric} GSO
projection may be used to evade this problem.

Since it is {\it a priori\/} unclear if complex-time generalizations of
quantum gravity and string theory can be formulated consistently,
we enclose the terms `spacetime' and `spacetime supersymmetry' in
quotation marks. It is one of our purposes to argue about possible consistency.

In this paper we continue to consider the critical $N{=}2$ string theory with
the global ${\bf Z}_2$ twist first introduced by Mathur and Mukhi \cite{mm}.
Our investigation can also be regarded as part of a more general search
for consistent flat backgrounds for $N{=}2$ string propagation, identifying
physical spectra and constructing corresponding scattering amplitudes.
As tools in our analysis, we use the BRST approach and
chiral bosonization~\cite{fms}. Since we are primarily interested in $N{=}2$
string scattering, we want any physical vertex operator to have a least
some continuous momenta  and, therefore, we ignore possible discrete states.

Employing the covariant bosonization lattice approach \cite{k-w} we list
all possible GSO projections in the ${\bf Z}_2$-twisted $N{=}2$ fermionic
string theory and show that they work in essentially the same way as in the
conventional NSR model. Namely, we identify two different chiral GSO-projected
$N{=}2$ string theories, one corresponding to the purely bosonic theory with
untwisted background  while another yielding a twisted local theory with bosons
and fermions. Each GSO restriction serves to project onto a {\it local} algebra
of BRST-invariant chiral vertex operators, therefore removing the major
obstruction against constructing interactions of twisted $N{=}2$ strings.

Like in the BRST-quantized
$N{=}1$ superstring, the superconformal ghost algebra in the $N{=}2$
case enforces an enlargement of the BRST cohomology by an infinite number of
copies of any physical state. This {\it picture} redundancy is made manifest
in the process of \pc. The complex nature of the superconformal
ghosts allows for two different bosonization schemes, resulting in a
{\it real\/} and a {\it holomorphic\/} version of \pc.
In contrast to the $N{=}1$ string, inverse \pc\ is not a {\it local\/}
operation. We explicitly construct non-local inverse \pco s in the real
version, which is required for the twisted model.

The ${\bf Z}_2$ twisting leads to a restriction of `spacetime' momenta
to $1{+}1$ dimensions, where massless kinematics essentially forbids
interactions.  Accordingly, we find that all twisted massless tree-level
three- and four-point functions vanish.  The untwisted theory, in distinction,
consists of a single {\it interacting\/} `spacetime' boson (and its partner
under spectral flow) at the massless level, as is well-known~\cite{ma}.
For definiteness, we restrict ourselves to the
{\it closed} $N{=}2$ string, but only write out the chiral half of it.

The paper is organized as follows. In sect.~2 we briefly review the well-known
case of the NSR model and the $N{=}1$ superstring, which are going to serve us
as the basic patterns. We then summarize the results of our previous work
\cite{klp} in sect.~3, which also introduces our notation and the two
bosonization schemes.  Sect.~4 is devoted to \pc\ in the $N{=}2$ string.
The fusion of chiral physical vertex operators is discussed in sect.~5.
In sect.~6 we introduce the covariant lattice approach which allows us
to identify the two GSO-projected models of the $N{=}2$ string.
Sect.~7 comprises our conclusion, with some remarks on tree-level correlation
functions.  An extensive list of chiral vertex operators can be found in
an Appendix.
\vglue.2in

\section{NSR model and $N{=}1$ superstring}

The standard two-dimensional (2d) action of the NSR model,
$$S\ =\ -\fracmm{1}{2\p}\int d^2\x\,\sqrt{h}\left[ h^{\a\b}\pa_{\a}X\cdot
\pa_{\b}X +i\bar{\j}\cdot\r^{\a}\pa_{\a}\j +(\pa_{\a}X+\bar{\j}\c_{\a})\cdot
\bar{\c}_{\b}\r^{\a}\r^{\b}\j\right]~,\eqno(2.1)$$
describes the minimal coupling of the $N{=}1$ matter $(X^{\m},\j^{\m})$ to 2d,
$N{=}1$ supergravity $(e^a_{\a},\c_{\a})$, and it is invariant under 2d
reparametrizations and local Lorentz rotations, $N{=}1$ supersymmetry, Weyl and
super-Weyl transformations.~\footnote{The dots stand for contractions of
target space indices.}  The latter are all non-anomalous in ${\bf R}^{1,9}$,
which
allows one to choose the gauge fixing $e^a_{\a}=\d^a_{\a},~\c_{\a}=0$. Then,
in light-cone coordinates after Wick rotation and gauge-fixing,
the action (2.1) takes the form
$$S\ =\ \fracmm{1}{\p}\int d^2 z\,\left[\pa X\cdot\bar{\pa} X -
 \ha\j\cdot\bar{\pa}\j + b\bar{\pa}c + \b\bar{\pa}\g\right] +
{\rm c.c.}~,\eqno(2.2)$$
where the usual conformal $(b,c)$ and superconformal $(\b,\g)$ ghosts
have been introduced.

The relative sign in boundary conditions for the 2d fermions and bosons is
well-known to be responsible for the two possible sectors (NS and R) of the
fermionic string. The 2d quantum
operators relating states in the NSR model are naturally divided into the
so-called {\it superfields\/} acting in each sector separately, and the
so-called {\it spin fields\/} intertwining the two sectors \cite{fms}.
There is an $N{=}1$
world-sheet supersymmetry in each of the sectors {\it separately},
while the full NSR theory is non-local. To express the spin fields $S^\a$
in terms of the NSR fermions $\j^\m$, one uses chiral
bosonization: $\j^{\m}\to e^{\pm\f^i}$ and $(\b,\g)\to e^{\pm\vf}$. This
introduces a $(5{+}1)$-dimensional half-integral weight lattice
$\L_w=(o)\cup(v)\cup(s)\cup(c)$, with the conjugacy classes $(v)$, $(s)$, and
$(c)$ being copies of a root lattice (conjugacy class $(o)$), shifted by an
elementary vector, spinor, or conjugate spinor weight, respectively,
just as for $so(8)$.~\footnote{
See, for example, refs.~\cite{kaku,myb} for a review.}
In order to get a local theory, one must project to an
{\it integral} sublattice of $\L_w$. This is accomplished by either one of two
GSO projections \cite{gos} which restrict the NSR model either to the
representations in $(o)\cup(v)$ or to those in $(o)\cup(s)$. \footnote
{Triality in $\L_w$ yields $(o)\cup(c)$ as a third possibility which is,
however, equivalent to $(o)\cup(s)$ by helicity flip or conjugation.}
The GSO projection of $\L_w$ to $(o)\cup(s)$
is known to give a spacetime supersymmetric theory --- the $N{=}1$
superstring \cite{gos}.
Beyond the tree level, the necessity for a GSO projection also derives from
the demand for modular invariance of superstring loop amplitudes, which is
ensured by the self-duality of the sublattice $(o)\cup(s)$.

The ten-dimensional spacetime supersymmetry generators are known to be just
the zero modes of the fermionic vertex operators at vanishing
momenta.~\footnote{
We ignore the trivial factor proportional to a fermionic wave
function.}
For the corresponding supercurrents one finds~\footnote{
The spinor indices are raised and lowered by use of the
ten-dimensional charge conjugation matrix.}
$$
Q^\a_{(-1/2)}(z)\ \sim\ e^{-{\frac 12}\vf}S^\a(z) \quad,\qquad
Q^\a_{(+1/2)}(z)\ \sim\ e^{+{\frac 12}\vf}(\pa X^\m)S_\b\g_\m^{\b\a}(z)
\quad,\eqno(2.3)$$
in the two different pictures indicated by a subscript in brackets.
The (affine) spacetime supersymmetry algebra takes the form
$$
Q^\a_{(-1/2)}(z)\; Q^\b_{(-1/2)}(w)\ \sim\ \fracmm{2}{z-w}\,
\g_\m^{\a\b}P^\m_{(-1)}(w)~,\eqno(2.4)$$
with the space-time translation generator $P^{\m}$ given by $e^{-\vf}\j^{\m}$
in the $(-1)$ picture, or just $\pa X^{\m}$ after
\pc\ to the $(0)$ picture. The currents in eqs.~(2.3) and (2.4)
form a closed $N{=}1$ spacetime supersymmetry algebra {\it modulo} \pc.
For example, the OPE~\cite{fms}
$$
P_{\m\,(0)}(z)\; Q^\a_{(+1/2)}(w)\ \sim\ \fracmm{1}{(z-w)^2}\,
e^{+{\frac 12}\vf} S_\b\g_\m^{\b\a}(w) \eqno(2.5)$$
does not formally vanish, as required by the super-Poincar\'e
algebra, but it picture-changes to zero instead.

There are some complications arising in the BRST analysis of conventional
$N{=}1$ superstrings, which have to be properly taken into account. First,
there
is a duplication of the BRST cohomology due to zero modes of the anticommuting
conformal ghosts, $\{b_0,c_0\}=1$, which are responsible for the
existence of {\it two\/} BRST-invariant operators $c V_1$ and $c\pa c V_1$ with
different ghost numbers for every ghost-independent physical vertex operator
$V_h$ of conformal dimension $h=1$. In addition, there are infinitely many
picture copies $V_{(p)}$, where $p\in {\bf Z}$ for the NS states and
$p\in {\bf Z}+\ha $ for the R states, due to the zero modes of the commuting
superconformal ghosts, $\[\b_0,\g_0\]=1$. The latter explains the need
for \pc~\cite{nst}.
\vglue.2in

\section{Twisting the $N{=}2$ string}

Our notation is as follows. Target space indices (internal and Lorentz) always
appear as superscripts.
The complex bosonic and fermionic matter fields are split either into real and
imaginary components,
$$Z^\m = Z^{2\m}+iZ^{3\m}\quad,\qquad
\J^\m = \J^{2\m}+i\J^{3\m}\qquad\qquad \m=0,1~,\eqno(3.1a)$$
or into holomorphic and antiholomorphic parts,
$$(Z^\m,Z^{\m*}) = (Z^{+\m},Z^{-\m})\quad,\qquad
(\J^\m,\J^{\m*}) = (\J^{+\m},\J^{-\m})\qquad\qquad \m=0,1~.\eqno(3.1b)$$
These fields may also be grouped into light-cone combinations~\footnote{
Compared to ref.~\cite{klp}, we have
slightly changed our notation here by using $\pm$ instead of $\ua\da$.}
$$\eqalign{
Z^{i\pm}=Z^{i0}\pm Z^{i1}\quad,&
\qquad\J^{i\pm}=\J^{i0}\pm \J^{i1}\qquad\qquad i=2,3 \cr
Z^{a\pm}=Z^{a0}\pm Z^{a1}\quad,&
\qquad\J^{a\pm}=\J^{a0}\pm \J^{a1}\qquad\qquad a=\pm~, \cr}\eqno(3.2)$$
with respect to the 2d Lorentz index $\m=0,1$.
Hence, lower-case Greek indices $\m,\n=0,1$ refer to a 2d Minkowski
space of signature $(-,+)$, while the lower-case Latin indices $i,j=2,3$
specify the real and imaginary components of the complex fields, and $a=\pm$
denote their holomorphic and antiholomorphic parts. To avoid confusing the
same numerical values of $\m$ and $i$, we have taken an unusual range for
the lower-case Latin indices, so that $\{i\m\}=\{20,21,30,31\}$.
In eq.~(3.2), on the other hand, the meaning of the signs must be inferred
from their position in the double $\pm$ index.
We frequently suppress inessential Lorentz indices.
For any two vectors $A^\m$ and $B^\n$, we thus have
$$A\cdot B\ \equiv\ \h_{\m\n}A^\m B^\n\ =\ -A^0B^0+A^1B^1\ =\
-\ha(A^+B^-+A^-B^+)~.\eqno(3.3)$$

The untwisted ({\it Ooguri-Vafa-type}) boundary conditions are \cite{ov}
$$Z(\p)\ =\ Z(0)\quad,\qquad
\J(\p)\ =\ ({+}\,{\rm or}\,{-})\J(0) \eqno(3.4)$$
jointly for all components.
They are the only ones which allow us to keep the single-valuedness of
the bosonic matter fields and, hence, yield ${\bf C}^{1,1}$ as the consistent
$(2{+}2)$-dimensional background `spacetime' for $N{=}2$ string propagation.
This choice of untwisted boundary conditions only deals with untwisted line
bundles and their square roots (spin bundles) to define fermions, just as
for the $N{=}1$ string. The two possible signs in eq.~(3.4) are common
for all the world-sheet spinors, and correspond to the usual NS-R distinction
familiar from the $N{=}1$ case. In other words, untwisted states reside in
the (NS,NS) or (R,R) sectors, where the two factors refer to the boundary
conditions of $(\J^2,\J^3)$.~\footnote{
Again our notation differs from ref.~\cite{klp}.}

The only ({\it Mathur-Mukhi-type}) example of twisted boundary conditions
that we employ in this paper is the one first considered in ref.~\cite{mm}:
$$Z^i(\p)\ =\ (-1)^i\,Z^i(0)\quad,\qquad
\J^i(\p)\ =\ ({+}\,{\rm or}\,{-})(-1)^i\,\J^i\qquad\qquad i=2,3~, \eqno(3.5a)$$
which in the holomorphic basis reads
$$Z^\pm(\p)\ =\ Z^\mp(0)\quad,\qquad
\J^\pm(\p)\ =\ ({+}\,{\rm or}\,{-})\J^\mp(0)~. \eqno(3.5b)$$
This means that fields of {\it integral} spin are allowed to pick up signs,
i.e. be {\it double}-valued, just as for fields of {\it half-integral} spin.
The complete monodromy behavior is now fixed by the signs picked up
by the components of $Z$, in addition to an overall sign between $Z$ and $\J$
related to the NS-R distinction. Twisted states comprise (NS,R) and (R,NS).
The twist implies the absence of the center-of-mass position and momentum
for $Z^3(\s)$, and it breaks $N{=}2$ to $N{=}1$ world-sheet supersymmetry.

The 2d (Brink-Schwarz) action for the $N{=}2$ fermionic string is well-known
\cite{bsa} (see also ref.~\cite{klp}).
The most important new features beyond the $N{=}1$ string action
are just the complexity of the 2d matter fields and the gravitino, and
the appearance of a (real) abelian gauge field (graviphoton) from the
world-sheet $N{=}2$ supergravity multiplet. The additional gauge fields are
responsible for more first-class constraints and the corresponding ghosts
after covariant gauge-fixing.

Both untwisted and twisted types of boundary conditions are blind to the
Lorentz indices of $Z$ and $\J$ and thus compatible with naive real
`spacetime' Lorentz symmetry~$O(1,1)$. The full global continuous symmetry of
the $N{=}2$ string action, however, is larger and  given by
$U(1,1) = \left[ U(1)\otimes SU(1,1)\right]/{\bf Z}_2$.

Interpreting the `internal' index~$i$ as part of a $(2{+}2)$-dimensional
`extended spacetime' label~$i\m$,
the extended spacetime Lorentz symmetry for the $N{=}2$ string should be
$SO(2,2)=\left[ SU(1,1)\otimes SU(1,1)\right]/{\bf Z}_2$ ~\cite{ket}. Although
the {\it gauge-fixed\/} $N{=}2$ string action has this symmetry (in the matter
part), the interaction terms in the {\it gauge-invariant\/} action break
one of the two $SU(1,1)$ factors down to $U(1)\otimes{\bf Z}_2$, with
${\bf Z}_2$ representing the Mathur-Mukhi twist. It has been conjectured
\cite{siegel} that the total global symmetry of the $N{=}2$ string  might
actually be the remnant of a `hidden' $SO(2,2)$ symmetry, which presumably
exhibits itself in the equivalent $N{=}4$ supersymmetric formulation of the
same theory. Anticipating this `extended Lorentz' symmetry, we are able to
distinguish between `bosons' and `fermions' in the target space, since
$SO(2,2)$ has representations of continuous spin and its little group
$GL(1)$ is non-trivial.~\footnote{The little group of $U(1,1)$ is trivial.}

Via BRST quantization in the $N{=}2$ superconformal gauge the fields (and
ghosts) of the $N{=}2$ string on the euclidean world-sheet become free, so
that they can be decomposed into their holomorphic and anti-holomorphic parts,
as is usual in 2d conformal field theory.  Lower-case~$\j$ will be used to
denote the chiral parts of~$\J$. In contrast,
the notation $Z$ is extended to the chiral part of the bosonic matter fields,
since $(z,\bar{z})$ is reserved for euclidean world-sheet coordinates.
The ghost systems appropriate for the $N{=}2$ string are:
\begin{itemize}
\item the reparametrization ghosts ($b,c$), an anticommuting pair of
free world-sheet fermions with conformal dimensions~($2,-1$).
\item the two-dimensional $N{=}2$ supersymmetry ghosts ($\b^i,\g^i$)
or ($\b^\mp,\g^\pm$), two commuting pairs of free world-sheet fermions
with conformal dimensions~($\frac32,-\frac12$).
\item the $U(1)$ ghosts ($\tilde{b},\tilde{c}$), an anticommuting pair of
free world-sheet fermions with conformal dimensions~($1,0$).
\end{itemize}

In order to construct fermionic vertex operators we make use of chiral
bosonization~\cite{fms,k-w}.
The reparametrization ghosts $(b,c)$ are expressed as
$$c\ \cong\ e^{+\s}~,\quad b\ \cong\ e^{-\s}~,
\qquad{\rm with}\quad \s(z)\ \s(w)\ \sim\ +\ln (z-w)~.\eqno(3.6)$$
Similarly, for the $U(1)$ ghosts $(\tilde{b},\tilde{c})$ one has
$$\tilde{c}\ \cong\ e^{+\tilde{\s}}~,\quad \tilde{b}\ \cong\ e^{-\tilde{\s}}~,
\qquad{\rm with}\quad \tilde{\s}(z)\ \tilde{\s}(w)\ \sim\ +\ln (z-w)~.
\eqno(3.7)$$
For the complex fields, bosonization depends on the basis. In the real basis,
$$\j^{i+}(z)\ \j^{j-}(w)\ \sim\ {2\ \d^{ij}\over z-w}\qquad{\rm and}\qquad
\g^i(z)\ \b^j(w)\ \sim\ {\d^{ij}\over z-w} \eqno(3.8a)$$
can be represented by~\footnote{
Normal ordering is always suppressed,
as well as cocycle operators~\cite{h,go,k-w}.}
$$\j^{i\pm}\ \cong\ \sqrt{2}\ e^{\pm\f^i} \quad,\qquad\qquad
\g^i\ \cong\ \h^i e^{\vf^i}\quad,\qquad \b^i\ \cong\ e^{-\vf^i}\pa\x^i~,
\eqno(3.9a)$$
$$\f^i(z) \f^j(w) \sim +\d^{ij}\ln (z-w)~,\quad
\vf^i(z) \vf^j(w) \sim -\d^{ij}\ln (z-w)~,\quad
\x^i(z) \h^j(w) \sim \fracmm{\d^{ij}}{z-w}~,$$
where four scalar bosons $\f^i$ and $\vf^i$ with values in $i({\bf R}/2\p)$
have been introduced.
The auxiliary anticommuting $(\h^i,\x^i)$ conformal system of spin (1,0)
may also be bosonized as
$$\x^i\ \cong\ e^{+\q^i}~,\quad \h^i\ \cong\ e^{-\q^i}~,\qquad{\rm with}\quad
\q^i(z)\ \q^j(w)\ \sim\ +\d^{ij}\ln (z-w)~.\eqno(3.10a)$$
Though the `solitons' $e^{\pm\vf^i}$ are outside the monomial field algebra of
$(\b^i,\g^i)$, one finds that \cite{ver}
$$e^{+\vf^i}\cong \d(\b^i)~,\quad e^{-\vf^i}\cong \d(\g^i)~,\quad
\x^i\cong \Theta(\b^i)~,\quad \h^i\cong \d(\g^i)\pa\g^i~.\eqno(3.11a)$$
Finally, we introduce real spin fields with helicity index $\pm$ as
$$S^{i\pm}\ \cong\ e^{\pm\frac12\f^i}~.\eqno(3.12a)$$
Their action on the vacuum state implements the Mathur-Mukhi twist for the
fermions and flips the boundary conditions of a given $\j^i$.

If one does not perform the Mathur-Mukhi twist, the holomorphic basis,
$$\j^{\pm+}(z)\ \j^{\mp-}(w)\ \sim\ \fracmm{4}{z-w} \quad,\qquad\qquad
\g^+(z)\ \b^-(w)\ \sim\ \fracmm{2}{z-w}~,\eqno(3.8b)$$
invites an alternative bosonization,
$$\j^{\pm+} \cong 2 e^{+\f^\pm}~,\qquad
\j^{\mp-} \cong 2 e^{-\f^\pm}~,\qquad\quad
\g^\pm \cong \h^\pm e^{+\vf^\pm}~,\qquad
\b^\mp \cong e^{-\vf^\pm}\pa\x^\mp~,\eqno(3.9b)$$
$$\f^\pm(z) \f^\pm(w) \sim +\ln (z-w)~,\quad
\vf^\pm(z) \vf^\pm(w) \sim -\ln (z-w)~,\quad
\x^\mp(z) \h^\pm(w) \sim \fracmm{2}{z-w}~.$$
Again, one may go on to
$$\x^\pm\ \cong\ \sqrt{2}\ e^{+\q^\pm}~,\quad
\h^\mp\ \cong\ \sqrt{2}\ e^{-\q^\pm}~,\qquad{\rm with}\quad
\q^\pm(z)\ \q^\pm(w)\ \sim\ +\ln(z-w)~.\eqno(3.10b)$$
For completeness, we mention that
$$e^{+\vf^\pm}\cong \d(\b^\mp)~,\quad e^{-\vf^\pm}\cong \d(\g^\pm)~,\quad
\x^\pm\cong \Theta(\b^\pm)~,\quad \h^\mp\cong \d(\g^\mp)\pa\g^\mp~.
\eqno(3.11b)$$
Finally, the holomorphic spin fields are
$$S^{\pm+}\ \cong\ e^{+\frac12\f^\pm} \qquad{\rm and}\qquad
S^{\mp-}\ \cong\ e^{-\frac12\f^\pm}~.\eqno(3.12b)$$
The two bosonization schemes are related by {\it non-local}
field redefinitions, as becomes clear by comparing, for example,
the expressions for the local matter $U(1)$ current,
$$\eqalign{
J\ &=\ \fracm{i}{2}\ \j^2\cdot\j^3\ =\
-\fracm{i}{2}\Bigl( e^{+\f^2-\f^3} + e^{-\f^2+\f^3} \Bigr) \cr
&=\ -\fracm14\ \j^+\cdot\j^-\ =\ \fracm12 (\pa\f^+-\pa\f^-)~.\cr}\eqno(3.13)$$
One reads off that $e^{q\f^\pm}$ has $U(1)$ charge $e{=}\pm q$.
Moreover, $\x^\pm$ and $\h^\pm$ are not linear combinations of
$\x^i$ and $\h^i$, as is the case for $\j$, $\g$ and $\b$.

Here and in what follows we use the standard results \cite{fms,k-w}
for free chiral bosons $\r\in\{\s,\tilde\s,\f,\vf,\q\}$ with a background
charge $\tilde{Q}_{\r}$,
$$h\left[e^{q\r}\right]\ =\ \fracmm{\e}{2}q(q-\tilde{Q})~,\qquad
\r(z)\ \r(w)\ \sim\ \e\ln(z-w)~,\qquad \e=\pm 1~,$$
$$\tilde{Q}_\s=3~,\qquad \tilde{Q}_{\tilde\s}=1~,\qquad \tilde{Q}_\f=0~,
\qquad \tilde{Q}_\vf=-2~,\qquad \tilde{Q}_\q=1~,\eqno(3.14)$$
where the factor $\e$ takes into account statistics.~\footnote{
$\e=-1$ for $\r=\vf$, and $\e=+1$ otherwise.}
It follows that the spin
fields $S$ twisting $\j$ and $\c$ have conformal dimension
$h{=}1/8$. Similarly, the fields $e^{-\frac12\vf}$ twisting $\b$ and $\g$
have conformal dimension $h{=}3/8$.

It is important to realize that the bosonization~(3.9)--(3.11) of the
superconformal ghosts enormously enlarges the set of local fields.
Indeed, for the real as well as for the holomorphic scheme,
the extended Fock space based on $\vf$ and $\q$ is a direct sum of
$({\bf Z}\times{\bf Z})$ copies (`pictures') of the original $(\b,\g)$
Fock space, each labelled by two picture numbers (either $\p_i$ or $\p_\pm$)
which are the eigenvalues of
$$\Pi_i\ =\ -\oint[\b^i\g^i+\h^i\x^i]\ =\ \oint[-\pa\vf^i+\pa\q^i]
\quad\qquad i=2,3 \eqno(3.15a)$$
or
$$\Pi_\pm\ =\ -\fracm12\oint[\b^\pm\g^\mp+\h^\pm\x^\mp]\
=\oint[-\pa\vf^\mp+\pa\q^\mp]~,\phantom{xxxxxxx} \eqno(3.15b)$$
respectively.
Obviously, any polynomial only in $\b$ and~$\g$ has $\p{=}0$,
whereas $e^{q_i\vf^i}$ carries $\p_i{=}q_i$
and $e^{q\vf^\mp}$ has $\p_\pm{=}q$.
A further subtlety is that the extended Fock space does not contain the
constant zero modes of $\x$, since only derivatives of these fields
appear in eq.~(3.9).

To describe the $N{=}2$ string Fock space of states, we define the formal
super-$SL(2,{\bf C})$-invariant vacuum state $\ket{0}$ in the untwisted NS
sector, with vanishing momentum, conformal dimension and local $U(1)$ charge.
States in the (R,R) sector may then be created by the joint action of
two spin fields, $S^2S^3$ or $S^+S^-$.
In order to generate the twisted sectors (R,NS) or (NS,R), one acts with a
single spin field, $S^2$ or $S^3$, respectively.
However, this is not sufficient. In ref.~\cite{klp} we also had to utilize
the ${\bf Z}_2$-{\it twist} fields $t^{i\m}(z)$, whose role is to flip the
boundary conditions for $Z^{i\m}$. The twist fields generically act as
$$t^{i\m}(z)\ \pa Z^{j\n}(w)\ \sim\ \fracmm{\h^{\m\n}\ \d^{ij}}{\sqrt{z-w}}
\,\tilde{t}(w)~,\eqno(3.16)$$
where $\tilde{t}$ is another (excited) twist field. Twist fields act
trivially on $\j$. We assume that all twist fields are Virasoro primaries.
The conformal dimension of $t^{i\m}$ is $h{=}1/16$, and their OPE reads
$$t^{i\m}(z)\ t^{j\n}(w)\ \sim\ -\fracmm{\h^{\m\n}\ \d^{ij}}{(z-w)^{1/8}}~.
\eqno(3.17)$$
We find it convenient to define the composites
$$t^i\ \equiv\ t^{i0}t^{i1}~~({\rm no~sum!})\quad,\qquad
t^i_+\ \equiv\ e^{\tilde{\s}/2}\, t^i~, \eqno(3.18)$$
where $t^i_+$ have vanishing conformal dimension.

The $N{=}2$ string BRST charge
$$Q_{\rm BRST}\ =\ \oint_0 \fracmm{dz}{2\p i}\,j_{\rm BRST}(z)\ \equiv\
\oint j_{\rm BRST} \eqno(3.19)$$
can be found, for example, in Appendix B of our previous paper \cite{klp}.
In its most convenient form the dimension-one BRST current reads
$$\eqalign{
j_{\rm BRST}\ =&\ c\hat{T} + bc\pa c +\g^2G +\g^3{\Bar G}+\tilde{c}\hat{J}\cr
&-(\g^2\g^2 + \g^3\g^3)b +2i(\g^2\pa\g^3 - \g^3\pa\g^2)\tilde{b}
+\fracm34 \pa[c(\b^2\g^2 +\b^3\g^3)]\cr}\eqno(3.20a)$$
in the real basis, or
$$\eqalign{
j_{\rm BRST}\ =&\ c\hat{T} + bc\pa c + \fracm12\g^-G^+ + \fracm12\g^+G^-
+\tilde{c}\hat{J}\cr
&- \g^+\g^-b + (\g^-\pa\g^+ -\g^+\pa\g^-)\tilde{b}
+\fracm38 \pa[c(\b^+\g^- +\b^-\g^+)]\cr}\eqno(3.20b)$$
in its holomorphic form.
Here, we use the notation
$$\hat{T}\ =\ T_{\rm tot}-T_{b,c}\qquad{\rm and}\qquad
\hat{J}\ =\ J_{\rm tot}-\pa(\tilde{b}c)~,\eqno(3.21)$$
where $T_{b,c}=-2b\pa c -(\pa b)c$, and we introduced the full (BRST-invariant)
stress tensor $T_{\rm tot}$ and the $U(1)$ current $J_{\rm tot}$ as
$$\eqalign{
T_{\rm tot}\ =\ \{Q_{\rm BRST},b\}\ &=\
T+T_{b,c} -\tilde{b}\pa\tilde{c}-\fracm32
(\b^2\pa\g^2+\b^3\pa\g^3)-\fracm12(\g^2\pa\b^2+\g^3\pa\b^3)\cr
&=\ T+T_{b,c} -\tilde{b}\pa\tilde{c}-\fracm34
(\b^+\pa\g^-+\b^-\pa\g^+)-\fracm14(\g^+\pa\b^-+\g^-\pa\b^+)~,\cr
J_{\rm tot}\ =\ \{Q_{\rm BRST},\tilde{b}\}\ &=\
J +\pa(\tilde{b}c)-\fracm{i}2(\b^2\g^3-\b^3\g^2)\cr
&=\ J +\pa(\tilde{b}c)+\fracm14(\b^+\g^--\b^-\g^+)~.\cr}
\eqno(3.22)$$
Here $T$, $G$ and $J$ are the $N{=}2$ string (matter) currents
without ghosts, {\it viz.}
$$\eqalign{
T\ &=\ -\ha\left(\pa Z^i\cdot\pa Z^i - \j^i\cdot\pa\j^i\right)\
=\ -\ha\pa Z^+\cdot\pa Z^-
+\fracm14\j^+\cdot\pa\j^-+\fracm14\j^-\cdot\pa\j^+~,\cr
G\ &=\ \d^{ij}\pa Z^i\cdot\j^j \qquad,\qquad\qquad\qquad\qquad
G^+\ =\ \pa Z^-\cdot\j^+ ~,\cr
{\Bar G}\ &=\ \ve^{ij}\pa Z^i\cdot\j^j \qquad,\qquad\qquad\qquad\qquad
G^-\ =\ \pa Z^+\cdot\j^- ~,\cr
J\ &=\ \fracm{i}{4}\ve^{ij}\j^i\cdot\j^j\
=\ \fracm{i}{2}\j^2\cdot\j^3 \
=\ -\fracm14 \j^+\cdot\j^-~.\cr} \eqno(3.23)$$

Given any vertex operator $V$ of the type $cV_1$, where $V_1$ is conformal
primary of dimension one and built without $(b,c)$ ghosts, one easily
verifies that
$$\left[ c\hat{T}+bc\pa c \right](z)\;V(w)\ \sim\ {\rm regular}~.\eqno(3.24)$$
Furthermore, if $V_1$ is of the type $V_{1,0}$, where
$V_{1,0}$ has vanishing local $U(1)$ charge and does not contain
$(\tilde b,\tilde c)$ ghosts, one also has
$$\tilde c \hat J(z)\;V(w)\ \sim\ {\rm regular}~.\eqno(3.25)$$
$h(V_{1,0}){=}1$ and $e(V_{1,0}){=}0$ are just the physical state conditions
of on-shell momentum and charge neutrality.
Twisted states are, however, not covered by this argument since they require
a twisting of $(\tilde b,\tilde c)$ by $e^{\tilde\s/2}$.
Replacing $cV_1$ by $c\pa cV_1$, or $V_{1,0}$ by $\tilde{c}V_{1,0}$, creates
additional BRST-closed states which are, however, BRST-exact when off-shell
or non-neutral.  Finally, if $V_{1,0}\in(NS,NS)$ has the form
$e^{-\vf^2-\vf^3}W$, with a neutral,
zero-dimensional, and ghost-free $W$ satisfying
$G(z)W(w)\sim O({1\over z-w})\sim\Bar G(z)W(w)$,
BRST invariance of $V$ is guaranteed.  For $(R,R)$ states, one needs
$V_{1,0}=e^{-\frac12\vf^2-\frac12\vf^3}W'$, with a ghost-free $W'$, and
$G(z)W'(w)\sim O({1\over\sqrt{z-w}})\sim\Bar G(z)W'(w)$ yields Dirac equations.
More generally, the bosonic string constraints
$T_{\rm tot}$ and $J_{\rm tot}$ annihilate a given physical state
$\ket{\rm phys}$ only in the generalized Siegel gauge~\footnote{
The fermionic constraints $G_{\rm tot}$ are visible in the canonical
pictures $\p=-1,-\frac12$.}
$$ b_0\ket{\rm phys}\ =\ \tilde{b}_0\ket{\rm phys}\ =\ 0~.\eqno(3.26)$$
The constraints associated with the local $U(1)$ symmetry are absent for
twisted states, where $\tilde b_0$ does not exist.

As representatives of the massless physical states
in the {\it twisted\/} $N{=}2$ string, we choose the
following BRST-invariant (and BRST non-trivial) chiral vertex operators
(multiple sign choices are correlated)
$$\eqalign{
(NS,NS)~:~&~\quad
\F^{\pm}\ =\ c\,e^{-\vf^2-\vf^3}\,e^{-\frac{i}{2}k^{2\mp} Z^{2\pm}}~,\cr
(R,R)~:~&~\quad
\U^{\pm}\ =\ c\,e^{-\frac12\vf^2-\frac12\vf^3}\,S^{2\pm}S^{3\pm}\,
e^{-\frac{i}{2}k^{2\mp} Z^{2\pm}}~,\cr
(NS,R)~:~&~\quad
\X^{\pm}\ =\ c\,e^{-\vf^2-\frac12\vf^3}\,S^{3\pm}t^3_+\,
e^{-\frac{i}{2}k^{2\mp} Z^{2\pm}}~,\cr
(R,NS)~:~&~\quad
\L^{\pm}\ =\ c\,e^{-\frac12\vf^2-\vf^3}\,S^{2\pm}t^3_+\,
e^{-\frac{i}{2}k^{2\mp} Z^{2\pm}}~.\cr}
\eqno(3.27)$$
The first four (untwisted) operators just represent the NS-type
Ooguri-Vafa boson $(\F)$ and its R-type bosonic partner $(\U)$, while
the other four (twisted) operators correspond to the Mathur-Mukhi chiral
fermions $(\X,\L)$. Since the abelian gauge field has been twisted,
no states are related by spectral flow.
The twisted string supports a merely $(1{+}1)$-dimensional
kinematics, i.e. $k^3{=}0$, which turns massless states into either
`spacetime' left-movers ($k^{2+}{=}0$, the upper sign choice) or right-movers
($k^{2-}{=}0$, the lower sign choice).
As was shown in ref.~\cite{klp}, all the chiral vertex operators introduced in
eq.~(3.27) belong to the $N{=}2$ string BRST cohomology.~\footnote{
In ref.~\cite{klp} more general momenta were allowed for $\F$ and $\U$,
to cover the untwisted case as well.  See also refs.~\cite{bien,gr} for more
about the (untwisted) $N{=}2$ string BRST cohomology.}

The untwisted string, in contrast, lives in $2{+}2$ dimensions so that
massless states only satisfy
$$-2\ k^+\cdot k^-\ =\ k^{++}k^{--} + k^{+-}k^{-+}\ =\ 0\quad. \eqno(3.28)$$
In this case, $N{=}2$ supersymmetry allows only $(NS,NS)$ and $(R,R)$ sectors,
and holomorphic representations of massless physical states are
$$\eqalign{
(NS,NS)~:~&~\quad
\F\ =\ c\,e^{-\vf^--\vf^+}\,e^{ik\cdot Z}~,\cr
(R,R)~:~&~\quad
\U\ =\ c\,e^{-\frac12\vf^--\frac12\vf^+}\,
\Bigl(k^{\pm-}S^{-+}S^{++}\pm ik^{\pm+}S^{+-}S^{--}\Bigr)\,e^{ik\cdot Z}~,\cr}
\eqno(3.29)$$
where the choice of sign for $\U$ is irrelevant since the two expressions
agree via eq.~(3.28), as long as no $k^{a\pm}$ vanishes.
Spectral flow~\cite{ov} identifies $\F$ with $\U$, leading us back to the
Ooguri-Vafa result of a single massless scalar boson.
Alternatively, a real set of representatives is obtained upon replacing
$e^{q(\vf^-+\vf^+)}$ by $e^{q(\vf^2+\vf^3)}$ and $S^{\mp\pm}S^{\pm\pm}$ by
$S^{2\pm}S^{3\pm}$ in eq.~(3.29), and dropping the relative factor of~$i$
there. It is noteworthy that the restriction of (3.29) to the twisted
kinematics (either $k^{\pm+}{=}0$ or $k^{\pm-}{=}0$) splits the BRST
cohomology classes to $\F^\pm$ and $\U^\pm$ of eq.~(3.27).

To compute scattering amplitudes with the BRST-invariant chiral vertex
operators introduced in eqs. (3.27) and~(3.29), we need to discuss the \pc\
operations and GSO projections in $N{=}2$ string theory.
\vglue.2in

\section{Picture-changing}

The BRST cohomology problem is simplified by identifying grading operators.
For the $N{=}2$ string, these are
\begin{itemize}
\item the total ghost charge \\
$U=-\oint[bc+\tilde b\tilde c+\b^2\g^2+\b^3\g^3]
=\oint[\pa\s+\pa\tilde\s-\pa\vf^2-\pa\vf^3] \\
\phantom{U}=-\oint[bc+\tilde b\tilde c+\fracm12\b^+\g^-+\fracm12\b^-\g^+]
=\oint[\pa\s+\pa\tilde\s-\pa\vf^+-\pa\vf^-]$
\item the picture charges $\ \Pi_i=-\oint[\b^i\g^i+\h^i\x^i]
=\oint[-\pa\vf^i+\pa\q^i]$\quad\qquad $i=2,3$ \\
\phantom{the picture char} or
$\ \Pi_\pm=-\fracm12\oint[\b^\pm\g^\mp+\h^\pm\x^\mp]
=\oint[-\pa\vf^\mp+\pa\q^\mp]$
\item the full bosonic constraints $\ T_{\rm tot}$ and $J_{\rm tot}$
\end{itemize}
of which only $U$ does not commute with $Q_{\rm BRST}$.
Accordingly, it suffices to separately investigate simultaneous eigenspaces
of the commuting set $\{U,\Pi,L_0^{\rm tot},J_0^{\rm tot}\}$, labelled
by $\{u,\pi,h,e\}$. From
$L_0^{\rm tot}=\{Q_{\rm BRST},b_0\}$ and
$J_0^{\rm tot}=\{Q_{\rm BRST},\tilde b_0\}$
it readily follows that non-trivial cohomology only exists
for $h=e=0$.~\footnote{
Note, however, that $J_0^{\rm tot}$ and $e$ are not defined for twisted
states.}
At this point, no restriction on the values of $u$ or $\p$ arises,
so that an infinity of (massless) physical vertex operators is anticipated.
Like in the $N{=}1$ string, however, BRST cohomology classes differing by
integral values of $u$ or $\p$ correspond to the same physical state, if
their spacetime properties agree. Hence, we should identify physical states
with equivalence classes of BRST cohomology classes under integral total ghost
and picture number changes. The task of this section is to exhibit the form of
these equivalence relations and derive a number of representative vertex
operators.

For the evaluation of string amplitudes, the collection (3.27) of
representatives is not sufficient. Consider a correlation function of
exponential operators $e^{q(k)\r}$ for one of the ghost systems listed in
eq.~(3.14). The tree-level ghost number selection rules~\footnote{
Since the constant zero modes of $\x$ are not part of the Fock space,
the $\q$ charges must sum to zero instead of one.}
$$ \sum_k q(k) \ =\ -\tilde Q \eqno(4.1)$$
yield, in particular,
$$ \sum_k \pi(k) \ =\ -2\qquad{\rm and}\qquad
\sum_k u(k)\ =\ 0 \eqno(4.2)$$
for nonzero tree-level correlations of vertex operators.
These constraints generally require the use of vertex
operators in several pictures and total ghost sectors.

Let us begin with the representatives of eq.~(3.27) which have
$u(\F,\U,\X,\L)=(-1,0,0,0)$; their pictures can be read off the
$\vf^i$ charges.
We are looking for a map to equivalent vertex operators which changes the
$u$ and/or $\p_i$ values.
For the simplest BRST class in our list, represented by $\F$ with
$(u,\p_2,\p_3)=(-1,-1,-1)$, the total
ghost number~$u$ may be increased from $-1$ to $0$, $0$, or $+1$,
without changing the pictures, upon
replacing the factor~$c$ by $c\pa c$, $\tilde cc$, or $\tilde cc\pa c$,
respectively. This reflects the vacuum degeneracy due to the anticommuting
ghost zero modes in the untwisted sector, and leaves the generalized
Siegel gauge. A way to lower $u$ to $-2$ is exchanging the factor~$c$ by a
chiral integration $\int\!dz$, delocalizing the resulting vertex~$V^{\int}$.
The other vertex operators mentioned above, $\U$, $\X$ and $\L$,
have `partners' in non-zero ghost numbers as well.
In general, however, there does not seem to exist a simple recipe
to derive them directly from one another and establish an explicit equivalence.
Nevertheless, from the argument just given we expect in a given picture that
localized untwisted vertex operators come in four types and
localized twisted ones in two.

None of the above changes the picture numbers~$\p_i$.
The existence of the doubly infinite set of pictures is related to the
superconformal ghost algebra.
Fortunately, we do not need to repeat the cohomology analysis for each picture,
since in this case an explicit equivalence relation is known. More precisely,
the so-called \pc\ operations~$X^i$
shift $\pi_j\to\pi_j+\d^i_j$ while commuting with $Q_{\rm BRST}$, $b_0$
and~$L_0^{\rm tot}$, and raising $u$ by one.
The latter fact makes the total ghost number impractical to label
picture-equivalence classes of vertex operators, so we introduce instead the
real picture-invariant combination
$$V_r\ =\ U-\P_2-\P_3\ =\ -\oint[bc+\tilde b\tilde c-\h^2\x^2-\h^3\x^3]\
=\ \oint[\pa\s+\pa\tilde\s-\pa\q^2-\pa\q^3]~,\eqno(4.3a)$$
with eigenvalues
$$v_r\ =\ u-\p_2-\p_3~. \eqno(4.4a)$$
The $v_r$-charges in a tree-level correlation function must sum to~$4$.
We speculate that BRST cohomology exists only for $v_r=1,2,2',3$ in the
untwisted sectors,~\footnote{
The two BRST classes with $v=2$ are being distinguished by using a prime.}
and for $v_r={3\over2},{5\over2}$ in the twisted ones,
with the indicated multiplicities, and omitting $V^{\int}$.

We display the picture of a vertex operator by parenthesized subscripts
and its $v_r$-charge by a parenthesized superscript, $V^{(v_r)}_{(\p_2,\p_3)}$.
Picture-changing in the real basis then proceeds as
$$\eqalign{
V_{(\pi_2+1,\pi_3)}\ =&\ [Q_{\rm BRST},\x^2 V_{(\pi_2,\pi_3)}\}\ =:\
X^2\cdot V_{(\pi_2,\pi_3)} \cr
V_{(\pi_2,\pi_3+1)}\ =&\ [Q_{\rm BRST},\x^3 V_{(\pi_2,\pi_3)}\}\ =:\
X^3\cdot V_{(\pi_2,\pi_3)} ~,\cr}\eqno(4.5)$$
where, in the absence of normal ordering between $\x^i$ and $V$, one defines
the real \pco s
$$ X^i(z)\ :=\ \{Q_{\rm BRST},\x^i(z)\}\qquad\qquad i=2,3~.\eqno(4.6a)$$
Analogously, in the holomorphic basis,
$$ X^\pm(z)\ :=\ \{Q_{\rm BRST},\x^\pm(z)\} \eqno(4.6b)$$
shifts $\p_+$ or $\p_-$ by one unit, while leaving the eigenvalues
$$v_h\ =\ u-\p_+-\p_- \eqno(4.4b)$$
of
$$V_h\ =\ U-\P_+-\P_-\ =\
-\oint[bc+\tilde b\tilde c-\fracm12\h^-\x^+-\fracm12\h^+\x^-]\
=\ \oint[\pa\s+\pa\tilde\s-\pa\q^+-\pa\q^-] \eqno(4.3b)$$
unchanged. Again, we conjecture that (apart from discrete states)
physical states reside in $v_h=1,2,2',3$ sectors only.

By construction, the $X$ are BRST invariant but {\it not} BRST trivial
due to the lack
of the zero modes of $\x$ in the bosonization formulae~(3.9)~\cite{fms}.
The \pco s just introduced take the following
explicit form:
$$\eqalign{
X^2\ =~&~c\pa\x^2 +e^{\vf^2}\left[ G + \fracm{i}{2}\tilde{c}\b^3
 + 2i(\pa\tilde{b})\g^3 + 4i\tilde{b}\pa\g^3\right]
+2e^{2\vf^2}b\pa\h^2 +\pa(e^{2\vf^2}b)\h^2 ~,\cr
X^3\ =~&~c\pa\x^3 +e^{\vf^3}\left[ {\Bar G} - \fracm{i}{2}\tilde{c}\b^2
 - 2i(\pa\tilde{b})\g^2 - 4i\tilde{b}\pa\g^2\right]
 + 2e^{2\vf^3}b\pa\h^3 +\pa(e^{2\vf^3}b)\h^3 ~,\cr}\eqno(4.7a)$$
$$\eqalign{
X^+\ =~&~c\pa\x^+ +e^{\vf^-}\left[ G^+
+2\pa\tilde{b}\g^+ +4\tilde{b}\pa\g^+ -2b\g^+ \right]~,\cr
X^-\ =~&~c\pa\x^- +e^{\vf^+}\left[ G^-
-2\pa\tilde{b}\g^- -4\tilde{b}\pa\g^- -2b\g^- \right] ~.\cr}\eqno(4.7b)$$
It is clear that $X^\pm$ are not just linear combinations of $X^i$.
The two types of \pco s differ in two respects.
The holomorphic version does not contain $e^{2\vf}$ terms, and it also lacks
any $\tilde{c}$ dependence. The latter means that $X^\pm$ commute with
$J_0^{\rm tot}$, whereas $X^i$ do so only modulo BRST-exact terms.
The dimensions of various fields appearing in the picture-changing operators
are compiled in Tables~I and II. One can easily check, in particular, that
$h[X]=0$ and $u[X]=1$.

\noindent{\sf Table I}.
The conformal dimensions~$h$ of the $N{=}2$ string world-sheet fields
and ghosts.
\vglue.1in
\noindent\begin{tabular}{c|cccccccccccc}
\hline
field & $\pa Z$ & $\j$ & $b$ & $c$ & $\tilde{b}$ & $\tilde{c}$ &
$\b$ & $\g$ & $\d(\b)$ & $\d(\g)$ & $\h$ & $\x$ \\
\hline
h & $1$ & $1/2$ & $2$ & $-1$ & $1$ & $0$ &
$3/2$ & $-1/2$ & $-3/2$ & $1/2$ & $1$ & $0$ \\
\hline
\end{tabular}
\vglue.2in

\noindent{\sf Table II}.
The dimensions~$h$ of some frequently appearing exponential operators.
\vglue.1in
\noindent\begin{tabular}{c|cc|cccccc|ccc}
\hline
field & $e^{\pm\frac12\f}$ & $e^{\pm\f}$ &
$e^{-2\vf}$ & $e^{-\vf}$ & $e^{-\frac12\vf}$ &
$e^{+\frac12\vf}$ & $e^{+\vf}$ & $e^{+2\vf}$ &
$e^{-2\q}$ & $e^{-\q}$ & $e^{+\q}$ \\
\hline
h & $1/8$ & $1/2$ & $0$ & $1/2$ & $3/8$ & $-5/8$ & $-3/2$ & $-4$ &
$3$ & $1$ & $0$ \\
\hline
\end{tabular}
\vglue.2in

Since \pc\ $X$ establishes an equivalence of cohomology classes,
its inverse $Y$ can only be well-defined modulo BRST-trivial terms and may,
like $X$ itself, possess a BRST-trivial nonzero kernel. We require in the
real form that
$$ [Q_{\rm BRST},Y^i]\ =\ 0 \qquad\qquad{\rm and}\qquad\qquad
Y^2(z)\;X^2(w)\ \sim\ 1\ \sim\ Y^3(z)\;X^3(w) \eqno(4.8)$$
but do not constrain the mixed products.
The quantum numbers of $Y^i$ are determined as
$(h,u,\p_j)=(0,-1,-\d_{ij})$. A simple ghost number analysis shows that this
leaves only a single candidate for each $\vf^i$ ghost charge value below $-1$.
In the case of $Y^3$, for instance, we are forced to write a
linear combination of
$$ Y^3_k\ =\ c\,\left(\g^2\right)^{k-2}\,\pa^{k-1}\x^3\ldots\pa^2\x^3\pa\x^3\,
e^{-k\vf^3} \qquad\quad k\ge2 \eqno(4.9)$$
which satisfy
$$ Y^3_k(z)\,X^3(w)\ \sim\ \d_{k2} + O(z{-}w) ~.\eqno(4.10)$$
The promising first term,
$$ Y^3_2\ =\ c\,\pa\x^3\,e^{-2\vf^3}~,\eqno(4.11)$$
is identical with the inverse picture-changing operator of the $N{=}1$ string,
but fails to be BRST invariant for the $N{=}2$ string.
Astoundingly, this failure can be corrected by adding an infinite series of
$Y^3_k$, with $k=4,6,8,\ldots$. In other words, the coefficients in
$$ Y^3\ =\ \sum_{{k=2\atop{k~\rm even}}}^\infty
\left(\prod_{\ell=1}^{k-1}\ell!\right)^{-1} Y^3_k \ \ =\ \
Y^3_2+{1\over12}Y^3_4+{1\over34560}Y^3_6+\ldots\eqno(4.12)$$
lead to a chain of cancellations among BRST commutators of successive terms.
With some effort, the formal series can be summed to the non-local expression
$$ Y^3(w)\ =\ -\sin \oint_w [\g^2\b^3-\g^3\b^2] \cdot Y^3_1(w) \ =\
i \sinh (2~{\rm ad}J_0^{\rm tot})\cdot Y^3_1(w) \eqno(4.13)$$
where we introduced
$$ Y^3_1\ =\ -c\,\x^2\,e^{-\vf^2-\vf^3}\ =\ c(\g^2)^{-1}\d(\g^3) \eqno(4.14)$$
and understand the action on $Y^3_1$ as a power series of iterated commutators.
Of course, a mirror image expression emerges for~$Y^2$.
Note that the $Y^i$ are pure ghost operators and do not contain
any matter fields. The formal sum in eq.~(4.12) is no longer contained in the
local bosonized field algebra. Thus, $Y^i\cdot V$ may not correspond to
a state even in the extended Fock
space, unless only a finite number of terms from eq.~(4.12) contribute.
Furthermore, the issue of BRST equivalence becomes intractable. There are
instances where $Y\cdot(X\cdot V)-V$ produces an infinite series which ought to
be BRST trivial. In our calculations, we have used $Y$ as a guide to new
representatives but always checked the results by reapplying~$X$.
A similar analysis in the holomorphic basis fails to produce any candidate
for~$Y^\pm$. Still, we suspect that some, necessarily non-local, inverse \pco s
exist in this case as well.

The massless vertex operators in eq.~(3.27) or (3.29) may be written as
$$\F\ \equiv\ V^{(1)}_{(-1,-1)}\quad,\quad
\U\ \equiv\ V^{(1)}_{(-\frac12,-\frac12)}\quad,\quad
\X\ \equiv\ V^{(\frac32)}_{(-1,-\frac12)}\quad,\quad
\L\ \equiv\ V^{(\frac32)}_{(-\frac12,-1)}~, \eqno(4.15)$$
with additional helicity indices in the twisted string.
In the remainder of this section we shall present some of their cousins
in other pictures, as needed later in tree amplitudes.
The principal tools are picture-changing (which changes $\p$ but not $v$)
and fusion, $V_1(z)V_2(w)\sim V_3(w)$ (which is additive in $\p$ and $v$),
to be discussed in the following section. To simplify the notation, we
suppress the ubiquitous factor $e^{ik\cdot Z}$.
The masslessness condition reads
$$k^2\cdot k^2+k^3\cdot k^3\ =\ -(k^{2+}k^{2-}+k^{3+}k^{3-})\
=\ -\fracm12(k^{++}k^{--}+k^{+-}k^{-+})\ =\ 0~.\eqno(4.16)$$
Since the untwisted vertex operators of the twisted model are obtained from
restricting the momenta of the real vertex operators of the untwisted string,
it is convenient to discuss the full $(2{+}2)$-dimensional vertex operators for
$\F$ and $\U$ in the real representation. Although picture-changing fixes
the relative normalization of the vertex operators, we have not paid
attention to it.

For the Ooguri-Vafa boson, $\F$, it is straightforward to work upwards in
pictures as we already know $V^{(v)}_{(-1,-1)}$ for $v=0,1,2,2',3$.
Some $v_r{=}1$ vertex operators are
$$\eqalign{
V^{(1)}_{(-1,-1)}\ &=\ c\,e^{-\vf^2-\vf^3} \cr
V^{(1)}_{(\ 0,-1)}\ &=\ c\,e^{-\vf^3}(k{\cdot}\j)+\g^2\,e^{-\vf^3}+
\frac{i}2\tilde cc\pa\x^3e^{-2\vf^3}\cr
V^{(1)}_{(-1,\ 0)}\ &=\ c\,e^{-\vf^2}[k{\cdot}\j]+\g^3\,e^{-\vf^2}-
\frac{i}2\tilde cc\pa\x^2e^{-2\vf^2}\cr
V^{(1)}_{(\ 0,\ 0)}\ &=\ c\left\{[k{\cdot}\pa Z]-
[k{\cdot}\j](k{\cdot}\j)\right\}-
\g^2[k{\cdot}\j]+\g^3(k{\cdot}\j)\cr} \eqno(4.17)$$
where we have abbreviated
$$ (k{\cdot}\j)\ \equiv\ i\d^{ij}\,k^i\cdot\j^j \qquad{\rm and}\qquad
[k{\cdot}\j]\ \equiv\ i\e^{ij}\,k^i\cdot\j^j~,\eqno(4.18)$$
and similarly for $(k{\cdot}\pa Z)$ and $[k{\cdot}\pa Z]$.
An almost complete list for $v=0,1,2,2',3$, real and holomorphic,
are listed in the Appendix.

Next, let us consider the spectral-flow partner $\U$ in the
pictures $\p_i=-\frac12$ and $-\frac32$.
Rather than solving the fermionic constraints in the canonical
$(-\frac12,-\frac12)$ picture, it proves more straightforward to begin with
the subcanonical $(-\frac32,-\frac32)$ picture.
Applying $X^2$ and $X^3$ one obtains for $v_r{=}1$
$$\eqalign{
V^{(1)\pm}_{(-\frac32,-\frac32)}\ &=\
c\,e^{-\frac32\vf^2-\frac32\vf^3}S^{2\mp}S^{3\mp}\cr
V^{(1)\pm}_{(-\frac12,-\frac32)}\ &=\
c\,e^{-\frac12\vf^2-\frac32\vf^3}
[k^{2\mp}S^{2\pm}S^{3\mp}-ik^{3\mp}S^{2\mp}S^{3\pm}]
+\frac{i}2\tilde cc\pa\x^3 e^{-\frac12\vf^2-\frac52\vf^3}
S^{2\mp}S^{3\mp}\cr
V^{(1)\pm}_{(-\frac32,-\frac12)}\ &=\
c\,e^{-\frac32\vf^2-\frac12\vf^3}
[k^{3\mp}S^{2\pm}S^{3\mp}+ik^{2\mp}S^{2\mp}S^{3\pm}]
-\frac{i}2\tilde cc\pa\x^2 e^{-\frac52\vf^2-\frac12\vf^3}
S^{2\mp}S^{3\mp}\cr
V^{(1)\pm}_{(-\frac12,-\frac12)}\ &=\
c\,e^{-\frac12\vf^2-\frac12\vf^3}
[k^{\pm\mp}S^{2\pm}S^{3\pm}\pm k^{\pm\pm}S^{2\mp}S^{3\mp}]~,\cr}\eqno(4.19)$$
where the last expression obtains after division by $k^{\mp\mp}$.
The two sign choices are BRST-equivalent, which may be checked for
$V_{(-\frac32,-\frac32)}$ but is obvious from
$$\eqalign{
k^{-\pm}\,V^{(1)+}_{(-\frac12,-\frac12)}\ &\propto\
+k^{-\pm}[k^{+-}S^{2+}S^{3+}+k^{++}S^{2-}S^{3-}] \cr
&=\ \pm k^{+\pm}[k^{-+}S^{2-}S^{3-}-k^{--}S^{2+}S^{3+}]\ \propto\
\pm k^{+\pm}\,V^{(1)-}_{(-\frac12,-\frac12)}~,\cr}\eqno(4.20)$$
as long as all $k^{a\pm}$ are generically non-zero.
Each \pc\ by $X^i$ essentially raises the $\vf^i$ charge and applies one of
two Dirac operators,
$\slash{k}\equiv\d^{ij}k^i\sdot\G^j$ or
$\Bar{\slash{k}}\equiv\ve^{ij}k^i\sdot\G^j$, via
$$\pa Z^i\cdot\j^j(z)\quad S^{2\a}S^{3\b}e^{ik\cdot Z}(w)\ \sim\
(z-w)^{-3/2}\,(-ik^i)\cdot(\G^j S^2S^3)^{\a\b}e^{ik\cdot Z}~.\eqno(4.21)$$
We may write, for example,
$$k^{\mp\mp}\,V^{(1)\pm}_{(-\frac12,-\frac12)}\ =\
c\,e^{-\frac12\vf^2-\frac12\vf^3}
\left( \slash{k}\Bar{\slash{k}} S^2S^3\right)^{\mp\mp}~,\eqno(4.22)$$
or use $\slash{k}^+\slash{k}^-=-2i\slash{k}\Bar{\slash{k}}$ for a holomorphic
description, with $\slash{k}^\pm\equiv k^\mp\sdot\G^\pm$.
This ensures automatically that $V^{(1)}_{(-\frac12,-\frac12)}$
satisfies two Dirac equations,
$$\slash{k}\,V^{(1)}_{(-\frac12,-\frac12)}\ =\ 0\ =\
\Bar{\slash{k}}\,V^{(1)}_{(-\frac12,-\frac12)}\quad\qquad{\rm or}\quad\qquad
\slash{k}^\pm\,V^{(1)}_{(-\frac12,-\frac12)}\ =\ 0~,\eqno(4.23)$$
since each $\slash{k}$ squares to zero on-shell.
Such is not required for the BRST invariance of $V_{(-\frac32,-\frac32)}$.
As a rule, lower pictures enlarge the image of $Q_{\rm BRST}$ but increase
its kernel as well.
Again, states in other $v_r$ sectors as well as holomorphic representatives
can be found in the Appendix.

In the twisted theory, we must restrict the kinematics to $k^{3+}=0=k^{3-}$
which, together with masslessness, implies either
$$ k^{++}=0=k^{-+} \qquad\qquad {\rm or}\qquad\qquad k^{-+}=0=k^{--}~.
\eqno(4.24)$$
Now the upper and lower sign choices in eq.~(4.19) are no longer equivalent,
since division by $k^{\mp\mp}$ is no longer admissible, so that
$$k^{a+}=0\qquad\Longrightarrow\qquad k^{++}\,V^{(1)-}_{(-\frac12,-\frac12)}=0
\qquad{\rm and}\qquad k^{--}\,V^{(1)+}_{(-\frac12,-\frac12)}
\propto S^{2+}S^{3+} \eqno(4.25)$$
and the opposite in case $k^{a-}{=}0$.
Hence, the two above momentum constraints lead to different BRST classes,
i.e. $\U^+$ and $\U^-$. The helicity index distinguishes $(1{+}1)$-dimensional
`spacetime' left-movers ($\U^+$ with $k^{2+}{=}0$)
and right-movers ($\U^-$ with $k^{2-}{=}0$), and it agrees with the
spinor helicities in the $(-\frac12,-\frac12)$ picture.

The two fermions in the twisted sector, $\X$ and $\L$, are related by
interchanging indices $2\leftrightarrow3$ (except for twist fields and
momenta). To save space, we only display
vertex operators for $\X^+$ in the pictures $\p_2=-1,0$ and
$\p_3=-\frac32,-\frac12$. Like before, $\X^-$ (and $\L^-$) are obtained by
helicity flip. For $v_r=\frac32$ one finds
$$\eqalign{
V^{(\frac32)+}_{(-1,-\frac32)}\ &=\
c\,e^{-\vf^2-\frac32\vf^3}S^{3-}t^3_+ \cr
V^{(\frac32)+}_{(\ 0,-\frac32)}\ &=\ \left\{ c(k{\cdot}\j)+\g^2\right\}\,
e^{-\frac32\vf^3}S^{3-}t^3_+ \cr
V^{(\frac32)+}_{(-1,-\frac12)}\ &=\
c\,e^{-\vf^2-\frac12\vf^3}S^{3+}t^3_+ \cr
V^{(\frac32)+}_{(\ 0,-\frac12)}\ &=\ \left\{ c(k{\cdot}\j)+\g^2\right\}\,
e^{-\frac12\vf^3}S^{3+}t^3_+ \cr} \eqno(4.26)$$
Vertex operators with $v_r=\frac12$ and $\frac52$ appear in the Appendix.

We close the section by remarking that the non-vanishing tree-level two-point
functions induce a natural pairing among the vertex operators, relating $V$ to
its conjugate,
$$V^{(v)\pm}_{(\p_2,\p_3)} \qquad\longleftrightarrow\qquad
V^{(4-v)\pm}_{(-2-\p_2,-2-\p_3)}~.\eqno(4.27)$$
The non-vanishing two-point functions are made exactly from these pairs.
\vglue.2in

\section{Vertex operator algebra}

In this section we turn to the fusion algebra of the localized chiral
vertex operators discussed so far.  The relevant information is
contained in their various OPEs. More precisely, the fusion of
two BRST cohomology classes obtains as the constant piece in the OPE
of any two representatives.
The resulting vertex operator may be BRST-trivial,
as must be the case for all singular terms in the OPE.
In the previous section we have learned that the BRST cohomology classes
are labelled by $(v,\p_2,\p_3)$ which behave additively under fusion.
Since the spectrum of non-trivial $v$-values is presumably bounded, the
fusion of two vertex operators should automatically give a trivial answer
if the sum of their $v$-charges exceeds that range.
Ultimately, we like to divide the vertex operator algebra by the picture
and ghost number equivalence and arrive at a fusion algebra for the physical
excitations. In order to retain a non-trivial result we must keep
$v_r\in\{1,\frac32,2,\frac52,3\}$ for localized operators.
Since we investigate massless states only, the momenta $k$ and $k'$ of the
fusing vertex operators are taken to satisfy $k{\cdot}k'=0$.

On the one hand, as was already noticed in ref.~\cite{klp}, interactions
between the Ooguri-Vafa boson $\F$ and the Mathur-Mukhi fermions $(\L,\X)$ have
to be forbidden, since the relevant OPEs are {\it non-local}.
At vanishing momenta we have, for example,
$$\eqalign{
\F(z)\;\L^{\pm}(w)\ &\sim\ (z-w)^{-1/2}\,c\pa c\,e^{-{\frac 32}\vf^2-2\vf^3}
e^{\pm\frac12\f^2}\,t^3_+~,\cr
\L^+(z)\;\L^-(w)\ &\sim\ (z-w)^{-1/2}\,c\pa c\,e^{-\vf^2-2\vf^3}
\tilde{c}~,\cr}\eqno(5.1)$$
and similarly for $\X^\pm$.

On the other hand, each of the three sets
$(\F^+,\F^-,\U^+,\U^-)$, $(\X^+,\L^+,\U^-)$ or $(\X^-,\L^-,\U^+)$
separately have {\it local} OPEs among themselves. This means,
in particular, that an interacting theory can make sense for
each of those sets seperately.
Eq.~(5.1) also shows that if we just want to disregard the Ooguri-Vafa
boson~$\F$ in favor of the new fermionic fields,
we should restrict ourselves to a definite helicity.
This gives us two fermionic generators to play with,
for example $\L^+$ and $\X^+$.
Their fusions must either be trivial or lead to some representatives of~$\U^-$.

Generally speaking,
the non-locality between  the `spacetime' bosonic and fermionic
vertex operators does not yet mean that their interactions are impossible,
since the left-moving (chiral) fields still have to be combined with the
right-moving ones to complete the full vertex operators. As the example of the
non-supersymmetric $O(16)\otimes O(16)$ string showed~\cite{nsu,dixhar,amp},
square root singularities in the OPEs may disappear when the proper GSO
projection is applied for modular invariance. We have investigated this
opportunity in some detail and found that an asymmetric GSO projection does
{\it not} work for the $N{=}2$ string, being unable to produce well-defined
correlation functions.

First, let us concentrate on the {\it bosonic} theory whose massless
physical spectrum is represented by $\F^\pm$ and~$\U^{\pm}$.
Allowing for the momenta to be $(2{+}2)$-dimensional amounts to
combining the four states to two, $\F$ and $\U$, which was shown in the
previous section.
A fusion of two $\F$ fields, as represented in eq.~(3.27),
with momenta $k$ and $k'$ and $k{\cdot}k'=0$ gives
$$\F(z)\;\F(w)\ \sim\
-(z{-}w)^{-1}\,c\pa c e^{-2\vf^2-2\vf^3} e^{i(k+k')\cdot Z}(w) +
\ha V^{(2)}_{(-2,-2)}(w) + O(z{-}w) ~,\eqno(5.2)$$
where the residue
$$c\pa c\,e^{-2\vf^2-2\vf^3} e^{i(k+k')\cdot Z}\ =\
\left\{ Q_{\rm BRST}\,,\,c\,e^{-2\vf^2-2\vf^3} e^{i(k+k')\cdot Z} \right\}
\eqno(5.3)$$
is BRST-trivial as expected, but the finite term
$$V^{(2)}_{(-2,-2)} =
\left\{ Q_{\rm BRST},\pa c\,e^{-2\vf^2-2\vf^3} e^{i(k+k')\cdot Z} \right\}
- c\pa c\,i(k{-}k'){\cdot}\pa Z\,e^{-2\vf^2-2\vf^3} e^{i(k+k')\cdot Z}
\eqno(5.4)$$
is not. In fact, consecutive application of $X^2$ and $X^3$ reproduces
$V^{(2)}_{(-1,-1)}$ of eq.~(A.1a). The same result may be observed in
other pictures, e.g. for
$$V^{(1)}_{(0,-1)}(z)\;V^{(1)}_{(-1,0)}(w)\ \sim\ V^{(2)}_{(-1,-1)}(w)~,
\eqno(5.5)$$
showing that fusion of $\F$ with itself reproduces $\F$.
For the fusion of $\U$ with $\F$ one finds with eq.~(4.22) that
$$\U(z)\;\F(w)\ \sim\ -c\pa c\,e^{-\frac32\vf^2-\frac32\vf^3}
\left(\slash{k}\Bar{\slash{k}}S^2S^3\right) e^{i(k+k')\cdot Z}(w)\
\propto\ V^{(2)}_{(-\frac32,-\frac32)}
\eqno(5.6)$$
which does not picture-change to zero since the momentum on the right-hand
side is $k{+}k'$.
Finally, fusion of $\U$ with itself yields
$$\U(z)\;\U(w)\ \sim\ -c\pa c\,e^{-\vf^2-\vf^3}\,
({\rm linear~in~}kk')\,e^{i(k+k')\cdot Z}(w)\
\propto\ V^{(2)}_{(-1,-1)}~.\eqno(5.7)$$

Denoting the picture and ghost number equivalence classes of cohomology classes
by square brackets, the physical fusion algebra in the bosonic (untwisted)
theory takes the simple form of
$$[\F]\cdot[\F]\ =\ [\F]\quad,\qquad
[\F]\cdot[\U]\ =\ [\U]\quad,\qquad
[\U]\cdot[\U]\ =\ [\F]\quad,\eqno(5.8)$$
which is consistent with the only non-zero three-point functions
$\VEV{\F\F\F}$ and $\VEV{\F\U\U}$.
The spectral flow identifies $\F$ with $\U$.
Indeed, a short calculation shows that,
at tree-level ({\it cf\/} ref.~\cite{mar}),
$$\VEV{\F\F\F}\ =\
\VEV{V^{(2)}_{(-1,-1)}\;V^{(1)}_{(\ 0,\ 0)}\;V^{(1)}_{(-1,-1)}}\ =\
k^2_2{\cdot}k^3_3-k^3_2{\cdot}k^2_3\ =\ \fracm{i}{2} c_{23} \eqno(5.9)$$
while
$$\VEV{\F\U\U}\ =\ \VEV{V^{(2)}_{(-1,-1)}\;V^{(1)}_{(-\frac12,-\frac12)}\;
V^{(1)}_{(-\frac12,-\frac12)}}\ =\
k^{--}_2 k^{++}_3 -k^{-+}_2 k^{+-}_3\ =\ c_{23} \eqno(5.10)$$
as well, with $c_{ij}\equiv k^+_i{\cdot}k^-_j -k^-_i{\cdot}k^+_j =-c_{ji}$
being the additional non-trivial $U(1,1)$ (but not $O(2,2)$) invariant.
The bosonic `scalar' $\F$ is known to represent the only excitation
of self-dual gravity, corresponding to deformations of the free K\"ahler
potential in $2{+}2$ dimensions~\cite{ov,ket}.

The alternative local algebra of physical vertex operators can be based
on $\X^+$ and $\L^+$ in the {\it twisted} theory.
Note that for this helicity choice only the $k^{2-}$ component of the
momenta $k^{i\m}$ stays non-zero.  The OPEs then read
$$\eqalign{
\X^+(z)\;\X^+(w)\ &\sim\
\tilde{c}c\pa c\,e^{-2\vf^2-\vf^3}\,e^{\f^3}
\,e^{-\frac{i}{2}(k^{2-}+k'^{2-})Z^{2+}}(w)
\quad \stackrel{X^2}{\longrightarrow}\ 0~,\cr
\L^+(z)\;\L^+(w)\ &\sim\
\tilde{c}c\pa c\,e^{-\vf^2-2\vf^3}\,e^{\f^2}
\,e^{-\frac{i}{2}(k^{2-}+k'^{2-})Z^{2+}}(w)
\quad \stackrel{X^3}{\longrightarrow}\ 0~,\cr
\X^+(z)\;\L^+(w)\ &\sim\
\tilde{c}c\pa c\,e^{-\frac32\vf^2-\frac32\vf^3}S^{2+}S^{3+}
\,e^{-\frac{i}{2}(k^{2-}+k'^{2-})Z^{2+}}(w)
\ =\ V^{(3)-}_{(-\frac32,-\frac32)}
\quad \stackrel{X^2X^3}{\longrightarrow}\ 0~,\cr} \eqno(5.11)$$
the last vertex being trivial due to its `wrong' helicity.
The first two fusion rules may also be obtained directly from
$$\eqalign{
V^{(\frac32)}_{(-1,-\frac12)}(z)\;V^{(\frac32)}_{(0,-\frac12)}(w)\ &\sim\ 0\cr
V^{(\frac32)}_{(-\frac12,-1)}(z)\;V^{(\frac32)}_{(-\frac12,0)}(w)\ &\sim\
0~.\cr}\eqno(5.12)$$
With $\U^-$ moving to the right and the fermions to the left, massless fusion
is impossible at non-zero momenta.
Finally,
$$\U^-(z)\;\U^-(w)\ \sim\ c\pa c\,e^{-\vf^2-\vf^3}
(k^{-+}k'^{--}+k^{--}k'^{-+}) e^{-\frac{i}{2}(k^{2-}+k'^{2-})Z^{2+}}(w)\
= 0 \eqno(5.12)$$
since $k^{a+}=0=k'^{a+}$.

Thus, the local fusion algebra of chiral massless vertex
operators in the twisted $N{=}2$ string theory becomes trivial
(away from $k{\equiv}0$).
Since there is no spectral flow in the twisted theory, $\U^-$ here
has nothing to do with the field $\F$ which is absent.
The fusion rules are consistent with the vanishing of the only
potentially non-trivial tree-level three-point function,
$$ \VEV{\X^+\L^+\U^-}\ =\ \VEV{V^{(\frac32)}_{(-1,-\frac12)}\;
V^{(\frac32)}_{(-\frac12,-1)}\;V^{(1)}_{(-\frac12,-\frac12)}}\ =\ 0~,
\eqno(5.13)$$
simply due to the massless $(1{+}1)$-dimensional kinematics,
for non-zero momenta.
An equivalent new theory can of course be obtained by
flipping the helicities of $(\X^+,\L^+,\U^-)$ to $(\X^-,\L^-,\U^+)$.
Obviously, the twisted string does not support interactions of its
massless excitations.

One may now try to build $N{=}2$ extended `spacetime supersymmetry' generators
from the two `spacetime' fermionic vertex operators at vanishing momenta.
Taking the integrated versions $V^{(\frac12)}_{(-\frac12,-1)}$ and
$V^{(\frac12)}_{(-1,-\frac12)}$ from eqs. (A.15) and~(A.11),
and replacing $\int$ by a contour integration, we arrive at
$$\eqalign{
Q^{\pm}_2\ \equiv\ \oint\,j^{\pm}_2(z)\ =\ \oint\, \left. \L^{\pm}\right|_{
c-{\rm omitted},~k=0}\ &\sim\
\oint\, e^{-{\frac 12}\vf^2}e^{-\vf^3}S^{2\pm}\,t^3_+~,\cr
Q^{\pm}_3\ \equiv\ \oint\,j^{\pm}_3(z)\ =\ \oint\, \left. \X^{\pm}\right|_{
c-{\rm omitted},~k=0}\ &\sim\
\oint\, e^{-\vf^2}e^{-{\frac 12}\vf^3}S^{3\pm}\,t^3_+~,
\cr}\eqno(5.15)$$
where $\oint\equiv \oint_0\, \fracmm{dz}{2\p i}$, and either $+$ or $-$
should be chosen. From eq.~(5.11) we have the OPEs
$$\eqalign{
j^{+}_2(z)\;j^{+}_2(w) ~&~\sim\ (z-w)^{-1}\,
\tilde c\,e^{-2\vf^3}\g_\m^{++} \j^{2\m} e^{-\vf^2} (w)~,\cr
j^{+}_3(z)\;j^{+}_3(w) ~&~\sim\ (z-w)^{-1}\,
\tilde c\,e^{-2\vf^2}\g_\m^{++} \j^{3\m} e^{-\vf^3} (w)~,\cr
j^{+}_2(z)\;j^{+}_3(w) ~&~\sim\ (z-w)^{-1}\,
\tilde c\,e^{-\frac32\vf^2-\frac32\vf^3} S^{2+}S^{3+} (w)~,\cr}\eqno(5.16)$$
and similarly for $j^{-}_2(z)\,j^{-}_3(w)$, whereas
$$j^{+}_2(z)\;j^{-}_2(w)\ \sim\ O(z-w)^{-3/2}\quad,\qquad
j^{+}_3(z)\;j^{-}_3(w)\ \sim\ O(z-w)^{-3/2}~.\eqno(5.17)$$
The operators $\j^{2\m}e^{-\vf^2}$ and $\j^{3\m}e^{-\vf^3}$
appearing on the r.h.s. of the first two OPEs in
eq.~(5.16) are the `spacetime' translation operators in the
$(-1)$~picture, just like in the $N{=}1$ superstring theory.
However, this does not ultimately come through since the $N{=}2$
picture-changing of the
{\it full} operators on the r.h.s. of eq.~(5.16) brings the first two to zero.
The third operator,
$$P_{23}\ \equiv\ \oint \tilde c\,e^{-\frac32\vf^2-\frac32\vf^3} S^{2+}S^{3+}~,
\eqno(5.18)$$
should represent $\U^-$. However,
picture-changing turns it into zero due to vanishing momenta.
This result is consistent with
$$\eqalign{
\Bigl\{V^{(\frac12)}_{(0,-\frac12)},V^{(\frac12)}_{(0,-\frac12)}\Bigr\}\ &
=0\cr
\Bigl\{V^{(\frac12)}_{(0,-\frac12)},V^{(\frac12)}_{(-\frac12,0)}\Bigr\}\ &
=0~,\cr}\eqno(5.19)$$
integrated vertex operators being taken from eqs. (A.12) and (A.16).
Therefore, instead of translation operators we get in fact zeros
on the r.h.s. of our `spacetime supersymmetry' algebra,
$$\left\{ Q^+_i,Q^+_j \right\}\ =\ 0 \qquad\qquad i,j=2,3~.\eqno(5.20)$$
These $N{=}2$ extended `supersymmetry' generators should
rather be interpreted as a kind of `exterior' derivatives.
\vglue.2in

\section{GSO projections}

Until now, our treatment of local operator algebras has not been systematic.
In this section finally we are going to employ a unified formalism which
simultaneously deals with all chiral vertex operators defined in arbitrary
pictures. For the purpose of mutual locality of vertex operators, we may
temporarily forget about their momentum-dependence~\footnote{
For the tree-level correlation functions, their momentum dependence is
essentially absorbed into the usual Koba-Nielsen factor.
In the twisted sector, the constrained kinematics (only $k^{2-}$ non-zero)
allows merely $Z^{2+}$ which does not get twisted.}
and concentrate on their ghost, spin, and twist field structure.
In operator products,
the $(b,c)$ and matter twist fields never lead to branch cuts,
since $t^3$ always occurs in combination with a ghost twist field, i.e.
$t^3_+=t^3 e^{\tilde\s/2}$, and those have meromorphic OPE with one another.
A difficulty may arise from the non-local operator product of these
combinations
with $(\tilde b,\tilde c)$ fields,
i.e. $e^{\tilde\s/2}(z)\,e^{\tilde\s}(w)\sim O(z-w)^{1/2}$.
For this reason, we at first restrict ourselves to vertex operators without
$(\tilde b,\tilde c)$ ghost content in the untwisted sectors.
This condition is merely technical and will turn out to be irrelevant later.
Now, only $\j^{i\pm}$ and $S^{i\pm}$ as well as $(\b^i,\g^i)$ and their spin
fields may ruin locality.  After real bosonization, eqs.~(3.9a)--(3.12a),
any vertex operator is a linear combination of terms proportional to~\footnote{
Eventual derivatives of $\f^i$ or $\vf^i$ are irrelevant again.}
$$\exp\left[p_2\f^2 + q_2\vf^2 + p_3\f^3 + q_3\vf^3\right]~,\eqno(6.1)$$
where $p_i$ and $q_i$ take integral or half-integral values.
Furthermore, the conformal dimension of~(6.1), $h=\ha\sum_i(p_i^2-q_i^2-2q_i)$,
must be an integer to enable BRST invariance.
Specifically, the $[p_2,q_2;p_3,q_3]$ values for the operators in eq.~(3.27)
are
$$\eqalign{
\F^{\pm}\ \to\ & \left[0,-1;0,-1\right]~,\cr
\U^{\pm}\ \to\ &  \left[\pm\ha,-\ha;\pm\ha,-\ha\right]~,\cr
\X^{\pm}\ \to\ &  \left[0,-1;\pm\ha,-\ha\right]~,\cr
\L^{\pm}\ \to\ &  \left[\pm\ha,-\ha;0,-1\right]~.\cr}\eqno(6.2)$$

Generally speaking, each pair $[p_i,q_i]$ belongs to
$\P^{1,1}\equiv{\bf Z}^{1,1}\cup\bigl[(\ha,\ha)+{\bf Z}^{1,1}\bigr]$
which may be regarded as a lorentzian weight lattice $\G_w$.
Here, the scalar product has been chosen as $[p,q]\cdot[p',q']=pp'-qq'$.
Like in the $N{=}1$ string and for $so(8)$,
the weight lattice decomposes into the root lattice $(o)$ and three copies
of it, each shifted by a different elementary weight vector and conventionally
denoted by $(v)$, $(s)$ and $(c)$.
Explicitly, these so-called {\it conjugacy classes}~$(r)$ are
$$\eqalign{
(o)~: &\qquad p,q\in {\bf Z}~,\quad p+q\in 2{\bf Z}~,\cr
(v)~: &\qquad p,q\in {\bf Z}~,\quad p+q\in 2{\bf Z}+1~,\cr
(s)~: &\qquad p,q\in {\bf Z}+\ha~,\quad p+q\in 2{\bf Z}+1~,\cr
(c)~: &\qquad p,q\in {\bf Z}+\ha~,\quad p+q\in 2{\bf Z}~,\cr}\eqno(6.3)$$
and each set~$(r)$ of weights~$[p,q]$ forms an equivalence class of
representations.
The classes $(o)$ and $(v)$ reside in the NS sector whereas $(s)$ and $(c)$
live in the R sector. In $1{+}1$ dimensions,
the lorentzian length-squared $p^2-q^2$ is an even integer except for weights
in~$(v)$ where it is odd.
On the other hand, the contribution to the conformal dimension~$h$ of the
operator~(6.1) is integral for~$(o)$ and
half-integral for $(v)$, $(s)$ and $(c)$.
The total weight lattice~$\G_w$ is half-integral, as is obvious from the
following table of lorentzian scalar products {\it modulo 1\/}
for any two weight vectors from $(r)$ and $(r')$.

\begin{center}
\noindent\begin{tabular}{c|cccc}
${\cdot}$ & (o) & (v) & (s) & (c) \\
\hline
(o) & 0 & 0 & 0 & 0 \\
(v) & 0 & 0 & $\frac12$ & $\frac12$ \\
(s) & 0 & $\frac12$ & 0 & $\frac12$ \\
(c) & 0 & $\frac12$ & $\frac12$ & 0 \\
\end{tabular}
\vglue.1in
\end{center}

\noindent
Combing two representations from $(r)$ and $(r')$ in a tensor product yields
representations from a single conjugacy class $(r+r')$ which is given below.

\begin{center}
\noindent\begin{tabular}{c|cccc}
${+}$ & (o) & (v) & (s) & (c) \\
\hline
(o) & (o) & (v) & (s) & (c) \\
(v) & (v) & (o) & (c) & (s) \\
(s) & (s) & (c) & (o) & (v) \\
(c) & (c) & (s) & (v) & (o) \\
\end{tabular}
\vglue.1in
\end{center}

For our purposes, we have to consider the combined weights
$[p_2,q_2;p_3,q_3]$ which form the still half-integral lattice
$\G=\G_w^{(2)}\oplus\G_w^{(3)}=\{(r_2;r_3)\}$.
In eq.~(3.27), we have representatives of 7 classes, as follows
$$\eqalign{
\F^\pm\in(v;v)\quad,&\qquad \U^+\in(c;c)\quad,\qquad \U^-\in(s;s)\quad,\cr
\X^+\in & (v;c)\quad,\qquad \L^+\in(c;v)\quad,\cr
\X^-\in & (v;s)\quad,\qquad \L^-\in(s;v)\quad.\cr}\eqno(6.4)$$
In total, there are $4^2=16$ conjugacy classes. It is natural to group them
into 8 {\it twisted\/} and 8 {\it untwisted\/} classes,
the former consisting of $(NS;R)$ and $(R;NS)$, and the latter formed by
$(NS;NS)$ and $(R;R)$.
Another coarse grading of the $2{+}2$ dimensional lattice~$\G$ consists of its
division into points of {\it even\/} and {\it odd\/} length-squared.
The latter comprise the 6 conjugacy classes~$(r_2;r_3)$ with a single~$(v)$;
the former live in the remaining 10 classes.
Collecting the contributions $h_i=\ha(p_i^2-q_i^2-2q_i)$ to the conformal
dimension, one sees that the 6 conjugacy classes containing a single~$(o)$
yield a half-integral result which rules them out for physical states.
The remaining 10 classes~$\G'$ do not form a lattice but neatly separate into
$$\eqalign{
{\rm even~untwisted:}\qquad & (o;o)\quad(v;v)\quad(s;s)\quad(c;c)
\quad(s;c)\quad(c;s) \cr
{\rm odd~twisted:}\qquad & (v;s)\quad(v;c)\quad(s;v)\quad(s;c)\quad.\cr}
\eqno(6.5)$$
Like in the $N{=}1$ string, we should like to identify commuting
(spacetime bosonic) vertex operators with {\it even untwisted\/} weights
and anticommuting (spacetime fermionic) vertex operators with
{\it odd twisted\/} weights.
The lattice consideration takes us out of the `canonical pictures' of
eq.~(3.27)
because the fusion algebra only closes in the infinite set of all pictures.
Since the $\q^i$ charges are always integral, the picture numbers $\p_i$ agree
{\it modulo 1\/} with~$q_i$, so that integral (half-integral) values of
$\p_2+\p_3$ coincide with even untwisted bosonic (odd twisted fermionic)
vertex operators, and we may use these terms interchangeably.

The relevance of the lattice description derives from the
basic OPE of two bosonized operators from eq.~(6.1),
$$\exp[p_i\f^i +q_i\vf^i](z)\,\exp[p'_i\f^i +q'_i\vf^i](w)
\sim (z-w)^{p_ip'_i - q_iq'_i} \exp[(p_i+p'_i)\f^i +(q_i+q'_i)\vf^i](w)~,
\eqno(6.5)$$
where the lorentzian signature in the pole order stems from the relative sign
between eqs. (3.6) and~(3.10).
This relates the mutual locality of two vertex operators to the integrality
of their weight's lorentzian scalar product.
Moreover, fusion simply corresponds to adding weights.
Essentially squaring the above multiplication table, one finds
for the known vertex operators the following table

\begin{center}
\noindent\begin{tabular}{c|ccccccc}
${\cdot}$ & $\F^\pm$ & $\U^+$ & $\U^-$ & $\X^+$ & $\X^-$ & $\L^+$ & $\L^-$ \\
\hline
$\F^\pm$ & $+$ & $+$ & $+$ & $-$ & $-$ & $-$ & $-$ \\
$\U^+$ & $+$ & $+$ & $+$ & $-$ & $+$ & $-$ & $+$ \\
$\U^-$ & $+$ & $+$ & $+$ & $+$ & $-$ & $+$ & $-$ \\
$\X^+$ & $-$ & $-$ & $+$ & $+$ & $-$ & $+$ & $+$ \\
$\X^-$ & $-$ & $+$ & $-$ & $-$ & $+$ & $+$ & $+$ \\
$\L^+$ & $-$ & $-$ & $+$ & $+$ & $+$ & $+$ & $-$ \\
$\L^-$ & $-$ & $+$ & $-$ & $+$ & $+$ & $-$ & $+$ \\
\end{tabular}
\vglue.1in
\end{center}

\noindent
where $`+'$ stands for locality and $`-'$ for non-locality.
It is obvious that the only closed mutually local subsets are
$(\F^+,\F^-,\U^+,\U^-)$, $(\X^+,\L^+,\U^-)$ and $(\X^-,\L^-,\U^+)$,
as stated before.

However, this analysis is not exhaustive.
Our task is to enumerate {\it all\/} local vertex operator subalgebras.
This amounts to classifying the {\it integral sublattices\/}
$\G_{\rm int}\subset\G'$.
Each such $\G_{\rm int}$ is obtained from $\G$ by some GSO projection and
leads to a different string model.
By inspection, one finds six maximal possibilities,
$$\eqalign{
I\qquad\qquad    & (o;o)\ \cup\ (v;v)\ \cup\ (s;s)\ \cup\ (c;c) \cr
II\qquad\qquad   & (o;o)\ \cup\ (v;v)\ \cup\ (s;c)\ \cup\ (c;s) \cr
III\qquad\qquad  & (o;o)\ \cup\ (s;s)\ \cup\ (v;c)\ \cup\ (c;v) \cr
IV\qquad\qquad   & (o;o)\ \cup\ (c;c)\ \cup\ (v;s)\ \cup\ (s;v) \cr
V\qquad\qquad    & (o;o)\ \cup\ (s;c)\ \cup\ (v;s)\ \cup\ (c;v) \cr
VI\qquad\qquad   & (o;o)\ \cup\ (c;s)\ \cup\ (v;c)\ \cup\ (s;v) ~.\cr}
\eqno(6.6)$$
By construction, these lattices are self-dual, which is expected to be crucial
for modular invariance.
Abbreviating $(o;o)\equiv(0)$ and the other three classes present at a time
by $(1),(2),(3)$, the fusion rules of all models can be unified to
$$
(0)\cdot(0)=(i)\cdot(i)=(0)\quad,\quad
(0)\cdot(i)=(i)\quad,\quad
(i)\cdot(j)=(k)\quad\qquad
\{i,j,k\}=\{1,2,3\}~,\eqno(6.7)$$
allowing for vanishing coefficients like in eq.~(5.11).
Hence, one may continue to project any of these six models further to
$(0)\cup(i)$ or even to~$(0)$.
Simultaneous helicity flip $(s)\leftrightarrow(c)$ in both factors $(r)$ and
$(r')$ relates the models $III$ and~$IV$ as well as $V$ and~$VI$.
One-sided helicity flips connects $I$ and~$II$ as well as $III$ through~$VI$.
Hence, there are only {\it two\/} types of essentially distinct
GSO projections, say, model~$I$, with four bosonic classes,
and model~$III$, with two bosonic and
two fermionic ones. But these theories have already been identified!
Namely, returning to eq.~(6.4) we see that model~$I$ is nothing but the
untwisted theory, whereas model~$III$ has the content of our
twisted theory. At this point, however, we established that
there are {\it no\/} further (maximal) options beyond those two.
It may be noted that models $II$, $V$ and $VI$ do not contain massless states
since no massless BRST-invariant $(c;s)$- or $(s;c)$-type ground state was
found in ref.~\cite{klp}. Thus, the three local sets of~\cite{klp}~
appear as models $I$, $III$ and $IV$.

The fact that each GSO projection gives us a weight {\it lattice\/} ensures
that fusion of vertex operators does not lead out of a given model.
However, we would also like to form equivalence classes with respect to
\pc. A glance at eqs. (3.20a) and (4.7a) shows that $Q_{\rm BRST}$
as well as $X^i$ are linear combinations of terms in $(o;o)$ and $(v;v)$.
Thus, we must expect part of their action to change the conjugacy class
of a given operator by adding $(v;v)$ to it according to
the fusion table above.
Indeed, $V^{(1)}_{(0,0)}$ in eq.~(4.17), for instance,
contains pieces in~$(o;o)$ while $V^{(1)}_{(-1,-1)}\equiv\F\in(v;v)$.
Hence, in the untwisted theory \pc\ mixes
$$(o;o)\ \leftrightarrow\ (v;v) \qquad\qquad{\rm and}\qquad\qquad
(s;s)\ \leftrightarrow\ (c;c) \quad. \eqno(6.8)$$

In contrast, the $(v;v)$ pieces of $Q_{\rm BRST}$ and $X^i$ seem to disturb
consistency with the {\it twisted\/} GSO projection, since $(v;v)$ is
{\it not\/} part of the twisted lattice (see model III of eq.~(6.6)).
Alas, just because of their relative non-locality those $(v;v)$ terms
never contribute to twisted vertex operators, so that the conjugacy class
of $\X$ or $\L$ is retained under \pc.
Still, the $\U$ operators are troublesome, since they exist both in the
twisted and untwisted model.
However, the `dangerous' parts of $G$ and $\Bar{G}$ in eq.~(4.7a) contain
$\pa Z^3$ which does not contract with $e^{ik\cdot Z}$ when $k^3{=}0$.
The remaining $(v;v)$ pieces of $X^i$ are just the $(\tilde{b},\tilde{c})$
ghost terms. Those produce $\tilde{c}$'s in various vertex operators, which
lead to an extra branch cut in the OPE with $t^3_+$ of
$\X^+$ or $\L^+$! However, all the `dangerous' terms fuse with twisted
vertex operators to $O(z-w)^{+1/2}\to0$, since they never occur for $v=1$
while non-zero fusion only happens for the following values of~$(v)$:
$$\eqalign{
&(1)\cdot(1)\ =\ (2)\qquad\qquad\qquad (1)\cdot(2')\ =\ (3)\cr
&(\frac32)\cdot(1)\ =\ (\frac52)\qquad\qquad\qquad
(\frac32)\cdot(\frac32)\ =\ (3)~.\cr} \eqno(6.9)$$
Hence, BRST cohomology seems to be consistent with the twisted GSO projection,
and the untwisted vertex operators in the extended Siegel gauge (no $\pa c$
or $\tilde c$) do set the general pattern, as was assumed in the beginning of
this section. Further evidence derives from the fact that (formally)
the inverse \pco s $Y^i\in(o;o)$.
Therefore, the lattice approch is also relevant for more
general vertex operators (containing $\tilde c$), and the twisted fusion
algebra is not affected.

{}From the analysis
of the type~III GSO projection in eq.~(6.6), we should expect a {\it second\/}
bosonic state in~$(o;o)$ in addition to $\U^-\in(s;s)$. Using picture
equivalence, this new state must be represented in the $(-1,-1)$ picture.
However, dimensional analysis easily shows that there can be no massless
$(o;o)$ state in this picture; the only massless state at all is~$\F\in(v;v)$!
We must conclude that for the twisted theory the $(o;o)$ class has no BRST
cohomology, i.e. contains only trivial states. This leaves us exactly with
the two fermions and one boson already known.

In summary, there are just two different GSO projections in $N{=}2$ string
theory, one leading to a purely bosonic model containing the
Ooguri-Vafa boson, while another giving non-interacting
bosons {\it and\/} fermions in the target space.
In analogy to the standard NSR model, where the `twisted' GSO projection
generates an $N{=}1$ spacetime supersymmetric theory,
one finds here an $N{=}2$  {\it extended} `spacetime supersymmetry'.
The latter is, however, itself `twisted' in the sense that the fermionic
generators are not the square roots of `spacetime' translations
but square to zero instead. This property may be
rather natural for the twisted $N{=}2$ fermionic string theory.
In particular, it does {\it not\/} require an equal number of
bosonic and fermionic physical degrees of freedom.
\vglue.2in

\section{Conclusion}

Requiring world-sheet locality for physical vertex operators, we used
the covariant lattice approach to establish that only {\it two\/}
distinct maximal GSO projections are possible for the ${\bf Z}_2$-twisted
$N{=}2$ fermionic string.
The first one simply retains the untwisted states which, on the
massless level, were proved to comprise two `spacetime' scalars,
to be identified via spectral flow.

The second projection should lead to an $N{=}2$ superstring,
yet it falls short in three respects.
First, the necessary `spacetime' twist restricts the kinematics from
$D{=}2{+}2$ to $D{=}1{+}1$, trivializing massless interactions.
Second, there is one bosonic state `missing' from the massless spectrum
of two `spacetime' fermions and one `spacetime' boson.
Third, the would-be `spacetime supersymmetry' generators are null operators
instead of producing `spacetime' translations.
Their geometrical significance is yet to be understood.
As is well-known, the Witten index~\cite{w} counts the mismatch of
bosonic and fermionic ground states. In our twisted $N{=}2$ string theory, this
index is non-vanishing, which implies that our $N{=}2$ extended
`spacetime supersymmetry' is unbroken.

The local massless vertex operator algebras for both GSO-projected $N{=}2$
string theories were extracted. No interactions occur in the twisted model,
whereas the untwisted bosons fuse in accordance with spectral flow.

Without twisting, a real and a holomorphic chiral bosonization scheme is
available, which lead to two different pairs of pictures and \pco s.
For the real version, we were able to explicitly construct non-local inverse
\pco s. This technology was then used to compute the massless vertex operators
for all ghost sectors in various pictures,
as needed in tree-level correlations.

An advantage of chiral bosonization is that all correlation
functions and OPEs are given in terms of {\it free} field expressions -- this
is usually called {\it vertex representation}~\cite{fk}. Chiral
bosonization also gave us a convenient way of handling the various spin fields.
We have recomputed all tree-level three-point functions
for the untwisted and the twisted model. As regards the four-point function,
it is known that
$$\VEV{\F\F\F\F}\ =\
\VEV{V^{(2')}_{(-1,-1)}\;V^{(0)}_{(-1,-1)}\;V^{(1)}_{(0,0)}\;V^{(1)}_{(0,0)}}
\eqno(7.1)$$
vanishes due to the special kinematics in $2{+}2$ dimensions, which yields
$$c_{12}c_{34}t+c_{23}c_{41}s-16stu\ =\ 0 \eqno(7.2)$$
for $c_{ij}\equiv k^+_i{\cdot}k^-_j -k^-_i{\cdot}k^+_j$
and $-4s_{ij}\equiv k^+_i{\cdot}k^-_j +k^-_i{\cdot}k^+_j$.
Because of spectral flow, the same should hold true for
$\VEV{\F\F\U\U}$ and $\VEV{\U\U\U\U}$.
Indeed, abbreviating $z_{ij}\equiv z_i-z_j$ and
$$\S^\pm_i\ \equiv\
\left( k^{\pm-}_i S^{-+}S^{++}\pm ik^{\pm+}_i S^{+-}S^{--} \right) (z_i)
\eqno(7.3)$$
and making use of $s+t+u=0$,
one computes, for example,
$$\eqalign{
\VEV{\U\U\U\U}\ =&\
\VEV{V^{(2')}_{(-\frac12,-\frac12)}\;V^{(0)}_{(-\frac12,-\frac12)}\;
V^{(1)}_{(-\frac12,-\frac12)}\;V^{(1)}_{(-\frac12,-\frac12)}} \cr
=&\ \int\!\!dz_2\;z_{13}z_{14}z_{34}\;\prod_{i<j}z_{ij}^{-2s_{ij}-\frac12}\;
\VEV{\S^+_1\;\S^-_2\;\S^+_3\;\S^-_4} \cr
=&\ -{\G(-2s)\G(-2t)\over\G(1-2u)}\left[\,
4(k^+_1{\cdot}k^-_2)(k^+_3{\cdot}k^-_4)\,t +
4(k^+_2{\cdot}k^-_3)(k^+_4{\cdot}k^-_1)\,s \right] \cr
=&\ 0 ~,\cr}\eqno(7.4)$$
because the expression in square brackets equals eq.~(7.2).
In fact, the vanishing of the four-point function is easiest to confirm
from the latter correlator due to the symmetry in the picture assignments.

It would be interesting to investigate consequences of our results for the
$N{=}2$ string field theory or string loop amplitudes.
This work is in progress.

\vglue.2in

\noindent{\bf Note added:}

\noindent
N. Berkovits and C. Vafa pointed out to us the relevance of their recent
work~\cite{bv2} where they imbed the $N{=}2$ string into a new $N{=}4$
topological string theory. This allows them to rewrite the critical
$N{=}2$ string $n$-point functions as correlators in the topological
theory, where their vanishing to all string loop orders can be proved
for any $n{>}3$. Furthermore, one has an $SU(1,1)$ worth of choices for
the complex conjugation ${\bf Z}_2$ which determines the imbedding.
This extra $SU(1,1)$ completes the global symmetry to $SO(2,2)$ as
anticipated in ref.~\cite{siegel}, but it is realized in a twistor sense.

\newpage

\noindent{\bf Acknowledgements}

\noindent
We acknowledge discussions with N.~Berkovits, S.~Samuel, J.H.~Schwarz
and C. Vafa.
S.V.K. would like to thank the theory groups of King's College, the
University of Maryland, and Harvard University for hospitality.
O.L. is grateful to the theory group of City College of CUNY for hospitality
extended to him.

\vfill

\noindent{\Large\bf Appendix: list of massless chiral vertex operators}

In this Appendix, we display almost all vertex operators for the massless
states $\F$, $\U^+$, $\X^+$ and $\L^+$ with $0\le v\le3$ in all pictures
$-2<\p\le0$. Untwisted states are given in the real as well as in the
holomorphic representation. The normalization of the vertex operators is left
undetermined. Like in section~4, the universal exponential factor
$e^{ik\cdot Z}$ will be suppressed, and $\U^+$ is given for unconstrained
momenta where it is BRST-equivalent to $\U^-$. The opposite helicities
follow from flipping all light-cone indices.

\newpage

We begin with the real representation. The $\F$ vertex operators are
$$\eqalign{ &(\p_2,\p_3)=(-1,-1):
\qquad\qquad\qquad\qquad\qquad\qquad\qquad\qquad\cr
v_r=0:&\quad  \int\,e^{-\vf^2-\vf^3} \cr
v_r=1:&\quad  c\,e^{-\vf^2-\vf^3} \cr
v_r=2:&\quad  c\pa c\,e^{-\vf^2-\vf^3} \cr
v_r=2':&\quad \tilde cc\,e^{-\vf^2-\vf^3} \cr
v_r=3:&\quad  \tilde cc\pa c\,e^{-\vf^2-\vf^3} \cr
}\eqno(A.1a)$$

$$\eqalign{ &(\p_2,\p_3)=(\ 0,-1):\cr
v_r=0:&\quad  \int\,\left\{e^{-\vf^3}(k{\cdot}\j)-
\frac{i}2\tilde c\pa\x^3e^{-2\vf^3}\right\} \cr
v_r=1:&\quad  c\,e^{-\vf^3}(k{\cdot}\j)+\g^2\,e^{-\vf^3}+
\frac{i}2\tilde cc\pa\x^3e^{-2\vf^3} \cr
v_r=2:&\quad  c\pa c\,e^{-\vf^3}(k{\cdot}\j) -
(\pa c\g^2{-}2c\pa\g^2)\,e^{-\vf^3} -
\frac{i}2\tilde cc\pa c\pa\x^3e^{-2\vf^3} \cr
v_r=2':&\quad \tilde cc\,e^{-\vf^3}(k{\cdot}\j)+
\tilde c\g^2\,e^{-\vf^3} +2ic\h^3 \cr
v_r=3:&\quad  \tilde cc\pa c\,e^{-\vf^3}(k{\cdot}\j) -
\tilde c(\pa c\g^2{-}2c\pa\g^2)\,e^{-\vf^3} -2ic\pa c\h^3 \cr
}\eqno(A.2a)$$

$$\eqalign{ &(\p_2,\p_3)=(-1,\ 0):\cr
v_r=0:&\quad  \int\,\left\{e^{-\vf^2}[k{\cdot}\j]+
\frac{i}2\tilde c\pa\x^2e^{-2\vf^2}\right\} \cr
v_r=1:&\quad  c\,e^{-\vf^2}[k{\cdot}\j]+\g^3\,e^{-\vf^2}-
\frac{i}2\tilde cc\pa\x^2e^{-2\vf^2} \cr
v_r=2:&\quad  c\pa c\,e^{-\vf^2}[k{\cdot}\j] -
(\pa c\g^3{-}2c\pa\g^3)\,e^{-\vf^2} +
\frac{i}2\tilde cc\pa c\pa\x^2e^{-2\vf^2} \cr
v_r=2':&\quad \tilde cc\,e^{-\vf^2}[k{\cdot}\j]+
\tilde c\g^3\,e^{-\vf^2} -2ic\h^2 \cr
v_r=3:&\quad  \tilde cc\pa c\,e^{-\vf^2}[k{\cdot}\j] -
\tilde c(\pa c\g^3{-}2c\pa\g^3)\,e^{-\vf^2} +2ic\pa c\h^2 \cr
}\eqno(A.3a)$$

$$\eqalign{ &(\p_2,\p_3)=(\ 0,\ 0):\cr
v_r=0:&\quad  \int\left\{ [k{\cdot}\pa Z]-[k{\cdot}\j](k{\cdot}\j)
+\ha(\g^2\b^3{-}\g^3\b^2)\right\} \cr
v_r=1:&\quad  c\Bigl\{ [k{\cdot}\pa Z]-[k{\cdot}\j](k{\cdot}\j)\Bigr\}-
\g^2[k{\cdot}\j]+\g^3(k{\cdot}\j) \cr
v_r=2:&\quad  c\pa c\Bigl\{ [k{\cdot}\pa Z]-[k{\cdot}\j](k{\cdot}\j)\Bigr\}+
(\g^2\pa\g^3{-}\g^3\pa\g^2)+ic\pa\tilde c\cr
&\qquad +(\pa c\g^2{-}2c\pa\g^2)[k{\cdot}\j]-
(\pa c\g^3{-}2c\pa\g^3)(k{\cdot}\j) \cr
}\eqno(A.4a)$$

\newpage

For $\U^+$ one finds, with lightlike $k\in{\bf C}^{1,1}$,
$$\eqalign{ &(\p_2,\p_3)=(-\fracm32,-\fracm32):
\qquad\qquad\qquad\qquad\qquad\qquad\qquad\qquad\cr
v_r=0:&\quad  \int\,e^{-\frac32\vf^2-\frac32\vf^3}S^{2-}S^{3-} \cr
v_r=1:&\quad  c\,e^{-\frac32\vf^2-\frac32\vf^3}S^{2-}S^{3-} \cr
v_r=2:&\quad  c\pa c\,e^{-\frac32\vf^2-\frac32\vf^3}S^{2-}S^{3-} \cr
v_r=2':&\quad \tilde cc\,e^{-\frac32\vf^2-\frac32\vf^3}S^{2-}S^{3-} \cr
v_r=3:&\quad  \tilde cc\pa c\,e^{-\frac32\vf^2-\frac32\vf^3}S^{2-}S^{3-} \cr
}\eqno(A.5a)$$

$$\eqalign{ &(\p_2,\p_3)=(-\fracm12,-\fracm32):\cr
v_r=0:&\quad  \int\,\left\{ e^{-\frac12\vf^2-\frac32\vf^3}
[k^{2-}S^{2+}S^{3-}-ik^{3-}S^{2-}S^{3+}]
+\frac{i}2\tilde c\pa\x^3 e^{-\frac12\vf^2-\frac52\vf^3}
S^{2-}S^{3-} \right\} \qquad\cr
v_r=1:&\quad  c\,e^{-\frac12\vf^2-\frac32\vf^3}
[k^{2-}S^{2+}S^{3-}-ik^{3-}S^{2-}S^{3+}]
+\frac{i}2\tilde cc\pa\x^3 e^{-\frac12\vf^2-\frac52\vf^3}
S^{2-}S^{3-} \cr
v_r=2:&\quad  c\pa c\,e^{-\frac12\vf^2-\frac32\vf^3}
[k^{2-}S^{2+}S^{3-}-ik^{3-}S^{2-}S^{3+}]\cr
&\qquad-c\h^2\,e^{+\frac12\vf^2-\frac32\vf^3}S^{2-}S^{3-}
+\frac{i}2\tilde cc\pa c\pa\x^3\,e^{-\frac12\vf^2-\frac52\vf^3}S^{2-}S^{3-} \cr
v_r=2':&\quad \tilde cc\,e^{-\frac12\vf^2-\frac32\vf^3}
[k^{2-}S^{2+}S^{3-}-ik^{3-}S^{2-}S^{3+}] \cr
v_r=3:&\quad  \tilde cc\pa c\,e^{-\frac12\vf^2-\frac32\vf^3}
[k^{2-}S^{2+}S^{3-}-ik^{3-}S^{2-}S^{3+}]
-\tilde cc\h^2\,e^{+\frac12\vf^2-\frac32\vf^3}S^{2-}S^{3-} \cr
}\eqno(A.6a)$$

$$\eqalign{ &(\p_2,\p_3)=(-\fracm32,-\fracm12):\cr
v_r=0:&\quad  \int\,\left\{ e^{-\frac32\vf^2-\frac12\vf^3}
[k^{3-}S^{2+}S^{3-}+ik^{2-}S^{2-}S^{3+}]
-\frac{i}2\tilde c\pa\x^2 e^{-\frac52\vf^2-\frac12\vf^3}
S^{2-}S^{3-}\right\} \qquad\cr
v_r=1:&\quad  c\,e^{-\frac32\vf^2-\frac12\vf^3}
[k^{3-}S^{2+}S^{3-}+ik^{2-}S^{2-}S^{3+}]
-\frac{i}2\tilde cc\pa\x^2 e^{-\frac52\vf^2-\frac12\vf^3}
S^{2-}S^{3-} \cr
v_r=2:&\quad  c\pa c\,e^{-\frac32\vf^2-\frac12\vf^3}
[k^{3-}S^{2+}S^{3-}+ik^{2-}S^{2-}S^{3+}]\cr
&\qquad-c\h^3\,e^{-\frac32\vf^2+\frac12\vf^3}S^{2-}S^{3-}
-\frac{i}2\tilde cc\pa c\pa\x^2\,e^{-\frac52\vf^2-\frac12\vf^3}S^{2-}S^{3-} \cr
v_r=2':&\quad \tilde cc\,e^{-\frac32\vf^2-\frac12\vf^3}
[k^{3-}S^{2+}S^{3-}+ik^{2-}S^{2-}S^{3+}] \cr
v_r=3:&\quad  \tilde cc\pa c\,e^{-\frac32\vf^2-\frac12\vf^3}
[k^{3-}S^{2+}S^{3-}+ik^{2-}S^{2-}S^{3+}]
-\tilde cc\h^3\,e^{-\frac32\vf^2+\frac12\vf^3}S^{2-}S^{3-} \cr
}\eqno(A.7a)$$

$$\eqalign{ &(\p_2,\p_3)=(-\fracm12,-\fracm12):\cr
v_r=0:&\quad  \int\,e^{-\frac12\vf^2-\frac12\vf^3}
[k^{+-}S^{2+}S^{3+} + k^{++}S^{2-}S^{3-}] \cr
v_r=1:&\quad  c\,e^{-\frac12\vf^2-\frac12\vf^3}
[k^{+-}S^{2+}S^{3+} + k^{++}S^{2-}S^{3-}] \cr
v_r=2:&\quad  c\pa c\,e^{-\frac12\vf^2-\frac12\vf^3}
[k^{+-}S^{2+}S^{3+} + k^{++}S^{2-}S^{3-}]\cr
&\qquad +c\h^2\,e^{+\frac12\vf^2-\frac12\vf^3}
[k^{3-}S^{2+}S^{3-}+ik^{2-}S^{2-}S^{3+}]/k^{--}\cr
&\qquad +c\h^3\,e^{-\frac12\vf^2+\frac12\vf^3}
[k^{2-}S^{2+}S^{3-}-ik^{3-}S^{2-}S^{3+}]/k^{--} \cr
v_r=2':&\quad \tilde cc\,e^{-\frac12\vf^2-\frac12\vf^3}
[k^{+-}S^{2+}S^{3+} + k^{++}S^{2-}S^{3-}] \cr
v_r=3:&\quad  \tilde cc\pa c\,e^{-\frac12\vf^2-\frac12\vf^3}
[k^{+-}S^{2+}S^{3+} + k^{++}S^{2-}S^{3-}]\cr
&\qquad +\tilde cc\h^2\,e^{+\frac12\vf^2-\frac12\vf^3}
[k^{3-}S^{2+}S^{3-}+ik^{2-}S^{2-}S^{3+}]/k^{--}\cr
&\qquad +\tilde cc\h^3\,e^{-\frac12\vf^2+\frac12\vf^3}
[k^{2-}S^{2+}S^{3-}-ik^{3-}S^{2-}S^{3+}]/k^{--} \cr
}\eqno(A.8a)$$

\newpage

The twisted states $\X^\pm$ are generated by
$$\eqalign{ &(\p_2,\p_3)=(-1,-\fracm32):
\qquad\qquad\qquad\qquad\qquad\qquad\qquad\qquad\cr
v_r=\fracm12:&\quad  \int\,e^{-\vf^2-\frac32\vf^3}S^{3-}t^3_+ \cr
v_r=\fracm32:&\quad  c\,e^{-\vf^2-\frac32\vf^3}S^{3-}t^3_+ \cr
v_r=\fracm52:&\quad  c\pa c\,e^{-\vf^2-\frac32\vf^3}S^{3-}t^3_+ \cr
}\eqno(A.9)$$

$$\eqalign{ &(\p_2,\p_3)=(\ 0,-\fracm32):
\qquad\qquad\qquad\qquad\qquad\qquad\qquad\qquad\cr
v_r=\fracm12:&\quad  \int\,(k{\cdot}\j)\,e^{-\frac32\vf^3}S^{3-}t^3_+ \cr
v_r=\fracm32:&\quad  \left\{ c(k{\cdot}\j)+\g^2\right\}\,
e^{-\frac32\vf^3}S^{3-}t^3_+ \cr
v_r=\fracm52:&\quad  c\pa c\,e^{-\frac32\vf^3}(k{\cdot}\j)S^{3-}
t^3_+ (\pa c\g^2 - 2c\pa\g^2)\,e^{-\frac32\vf^3}S^{3-}t^3_+\cr
&\qquad+\frac{i}{2}c\pa c\pa\x^3\,e^{-\frac52\vf^3}S^{3-}
e^{\frac32\tilde{\s}}t^3 \cr
}\eqno(A.10)$$

$$\eqalign{ &(\p_2,\p_3)=(-1,-\fracm12):
\qquad\qquad\qquad\qquad\qquad\qquad\qquad\qquad\cr
v_r=\fracm12:&\quad  \int\,e^{-\vf^2-\frac12\vf^3}S^{3+}t^3_+ \cr
v_r=\fracm32:&\quad  c\,e^{-\vf^2-\frac12\vf^3}S^{3+}t^3_+ \cr
v_r=\fracm52:&\quad  \fracmm{i}{\sqrt{2}}
c\pa c\,e^{-\vf^2-\frac12\vf^3} k^{2-}S^{3+}t^3_+
-c\h^3\,e^{-\vf^2+\frac12\vf^3}S^{3-}t^3_+ \cr
}\eqno(A.11)$$

$$\eqalign{ &(\p_2,\p_3)=(\ 0,-\fracm12):\cr
v_r=\fracm12:&\quad  \int\,c(k{\cdot}\j)\,e^{-\frac12\vf^3}S^{3+}t^3_+ \cr
v_r=\fracm32:&\quad  \left\{ c(k{\cdot}\j)+\g^2\right\}\,
e^{-\frac12\vf^3}S^{3+}t^3_+ \cr
v_r=\fracm52:&\quad
\fracmm{i}{\sqrt{2}}c\pa c\,e^{-\frac12\vf^3}(k{\cdot}\j)k^{2-}S^{3+}t^3_+
+\fracmm{i}{\sqrt{2}}(\pa c\g^2-2c\pa\g^2)\,e^{-\frac12\vf^3}k^{2-}S^{3+}t^3_+
\cr &\qquad
-\left\{c(k{\cdot}\j)+\g^2\right\}\h^3\,e^{\frac12\vf^3}S^{3-}t^3_+
-\fracm{i}{2}c\,e^{-\frac12\vf^3}S^{3-}e^{\frac32\tilde{\s}}t^3 \cr
&\qquad+ 3ic\h^3\pa\h^3\,e^{\frac32\vf^3}S^{3-}e^{-\frac12\tilde{\s}}t^3
+\sqrt{2}c\pa c\h^3\,e^{\frac12\vf^3}k^{2-}S^{3+}e^{-\frac12\tilde{\s}}t^3 \cr
}\eqno(A.12)$$

\newpage

Interchanging $2\leftrightarrow3$ except for $t^3_+$ and $k^i$,
the states $\L$ obtain as
$$\eqalign{ &(\p_2,\p_3)=(-\fracm32,-1):
\qquad\qquad\qquad\qquad\qquad\qquad\qquad\qquad\cr
v_r=\fracm12:&\quad  \int\,e^{-\frac32\vf^2-\vf^3}S^{2-}t^3_+ \cr
v_r=\fracm32:&\quad  c\,e^{-\frac32\vf^2-\vf^3}S^{2-}t^3_+ \cr
v_r=\fracm52:&\quad  c\pa c\,e^{-\frac32\vf^2-\vf^3}S^{2-}t^3_+ \cr
}\eqno(A.13)$$

$$\eqalign{ &(\p_2,\p_3)=(-\fracm32,\ 0):
\qquad\qquad\qquad\qquad\qquad\qquad\qquad\qquad\cr
v_r=\fracm12:&\quad  \int\,[k{\cdot}\j]\,e^{-\frac32\vf^2}S^{2-}t^3_+ \cr
v_r=\fracm32:&\quad  \left\{ c[k{\cdot}\j]+\g^3\right\}\,
e^{-\frac32\vf^2}S^{2-}t^3_+ \cr
v_r=\fracm52:&\quad  c\pa c\,e^{-\frac32\vf^2}[k{\cdot}\j]S^{2-}
t^3_+ (\pa c\g^3 - 2c\pa\g^3)\,e^{-\frac32\vf^2}S^{2-}t^3_+ \cr
&\qquad+\frac{i}{2}c\pa c\pa\x^2\,e^{-\frac52\vf^2}S^{2-}
e^{\frac32\tilde{\s}}t^3\cr
}\eqno(A.14)$$

$$\eqalign{ &(\p_2,\p_3)=(-\fracm12,-1):
\qquad\qquad\qquad\qquad\qquad\qquad\qquad\qquad\cr
v_r=\fracm12:&\quad  \int\,e^{-\frac12\vf^2-\vf^3}S^{2+}t^3_+ \cr
v_r=\fracm32:&\quad  c\,e^{-\frac12\vf^2-\vf^3}S^{2+}t^3_+ \cr
v_r=\fracm52:&\quad  \fracmm{i}{\sqrt{2}}
c\pa c\,e^{-\frac12\vf^2-\vf^3} k^{2-}S^{2+}t^3_+
-c\h^2\,e^{+\frac12\vf^2-\vf^3}S^{2-}t^3_+ \cr
}\eqno(A.15)$$

$$\eqalign{ &(\p_2,\p_3)=(-\fracm12,\ 0):\cr
v_r=\fracm12:&\quad  \int\,c[k{\cdot}\j]\,e^{-\frac12\vf^2}S^{2+}t^3_+ \cr
v_r=\fracm32:&\quad  \left\{ c[k{\cdot}\j]+\g^3\right\}\,
e^{-\frac12\vf^2}S^{2+}t^3_+ \cr
v_r=\fracm52:&\quad
\fracmm{i}{\sqrt{2}}c\pa c\,e^{-\frac12\vf^2}[k{\cdot}\j]k^{2-}S^{2+}t^3_+
+\fracmm{i}{\sqrt{2}}(\pa c\g^3-2c\pa\g^3)\,e^{-\frac12\vf^2}k^{2-}S^{2+}t^3_+
\cr &\qquad
-\left\{c[k{\cdot}\j]+\g^3\right\}\h^2\,e^{\frac12\vf^2}S^{2-}t^3_+
-\fracm{i}{2}c\,e^{-\frac12\vf^2}S^{2-}e^{\frac32\tilde{\s}}t^3 \cr
&\qquad + 3ic\h^2\pa\h^2\,e^{\frac32\vf^2}S^{2-}e^{-\frac12\tilde{\s}}t^3
+\sqrt{2}c\pa c\h^2\,e^{\frac12\vf^2}k^{2-}S^{2+}e^{-\frac12\tilde{\s}}t^3 \cr
}\eqno(A.16)$$

\newpage

Untwisted states can be represented in the holomorphic basis.
$\F$ is created by
$$\eqalign{ &(\p_+,\p_-)=(-1,-1):
\qquad\qquad\qquad\qquad\qquad\qquad\qquad\qquad\cr
v_h=0:&\quad  \int\,e^{-\vf^--\vf^+} \cr
v_h=1:&\quad  c\,e^{-\vf^--\vf^+} \cr
v_h=2:&\quad  c\pa c\,e^{-\vf^--\vf^+} \cr
v_h=2':&\quad \tilde cc\,e^{-\vf^--\vf^+} \cr
v_h=3:&\quad  \tilde cc\pa c\,e^{-\vf^--\vf^+} \cr
}\eqno(A.1b)$$

$$\eqalign{ &(\p_+,\p_-)=(\ 0,-1):
\qquad\qquad\qquad\qquad\qquad\qquad\qquad\qquad\cr
v_h=0:&\quad  \int\,k^-{\cdot}\j^+\,e^{-\vf^+} \cr
v_h=1:&\quad  c\,k^-{\cdot}\j^+\,e^{-\vf^+} \cr
v_h=2:&\quad  c\pa c\,k^-{\cdot}\j^+\,e^{-\vf^+} -2ic\h^+ \cr
v_h=2':&\quad \tilde{c}c\,k^-{\cdot}\j^+\,e^{-\vf^+}-2ic\h^+ \cr
v_h=3:&\quad  \tilde{c}c\pa c\,k^-{\cdot}\j^+\,e^{-\vf^+}
+2i\tilde{c}c\h^+ -2ic\pa c\h^+ \cr
}\eqno(A.2b)$$

$$\eqalign{ &(\p_+,\p_-)=(-1,\ 0):
\qquad\qquad\qquad\qquad\qquad\qquad\qquad\qquad\cr
v_h=0:&\quad  \int\,k^+{\cdot}\j^-\,e^{-\vf^-} \cr
v_h=1:&\quad  c\,k^+{\cdot}\j^-\,e^{-\vf^-} \cr
v_h=2:&\quad  c\pa c\,k^+{\cdot}\j^-\,e^{-\vf^-} -2ic\h^- \cr
v_h=2':&\quad \tilde{c}c\,k^+{\cdot}\j^-\,e^{-\vf^-}-2ic\h^- \cr
v_h=3:&\quad  \tilde{c}c\pa c\,k^+{\cdot}\j^-\,e^{-\vf^-}
+2i\tilde{c}c\h^- -2ic\pa c\h^- \cr
}\eqno(A.3b)$$

$$\eqalign{ &(\p_+,\p_-)=(\ 0,\ 0):
\qquad\qquad\qquad\qquad\qquad\qquad\qquad\qquad\cr
v_h=0:&\quad  \int\,\Bigl\{ (k^+{\cdot}\pa Z^- -k^-{\cdot}\pa Z^+)
-i(k^+{\cdot}\j^-)(k^-{\cdot}\j^+)\Bigr\} \cr
v_h=1:&\quad  c\,\Bigl\{ (k^+{\cdot}\pa Z^- -k^-{\cdot}\pa Z^+)
-i(k^+{\cdot}\j^-)(k^-{\cdot}\j^+)\Bigr\}
-(\g^+\,k^+{\cdot}\j^- -\g^-\,k^-{\cdot}\j^+) \cr
v_h=2:&\quad  (c\pa c+\g^+\g^-)\,\Bigl\{(k^+{\cdot}\pa Z^--k^-{\cdot}\pa Z^+)
-i(k^+{\cdot}\j^-)(k^-{\cdot}\j^+)\Bigr\}\cr
&\qquad +(\pa c+ick{\cdot}\pa Z+c\pa)\,
(\g^+k^+{\cdot}\j^--\g^-k^-{\cdot}\j^+) \cr
}\eqno(A.4b)$$

\newpage

The holomorphic vertex operators for $\U^+$ read
(two correlated sign choices are equivalent)
$$\eqalign{ &(\p_+,\p_-)=(-\fracm32,-\fracm32):
\qquad\qquad\qquad\qquad\qquad\qquad\qquad\qquad\cr
v_h=0:&\quad  \int\,e^{-\frac32\vf^--\frac32\vf^+}S^{+-}S^{--} \cr
v_h=1:&\quad  c\,e^{-\frac32\vf^--\frac32\vf^+}S^{+-}S^{--} \cr
v_h=2:&\quad  c\pa c\,e^{-\frac32\vf^--\frac32\vf^+}S^{+-}S^{--} \cr
v_h=2':&\quad \tilde cc\,e^{-\frac32\vf^--\frac32\vf^+}S^{+-}S^{--} \cr
v_h=3:&\quad  \tilde cc\pa c\,e^{-\frac32\vf^--\frac32\vf^+}S^{+-}S^{--} \cr
}\eqno(A.5b)$$

$$\eqalign{ &(\p_+,\p_-)=(-\fracm12,-\fracm32):
\qquad\qquad\qquad\qquad\qquad\qquad\qquad\qquad\cr
v_h=0:&\quad  \int\,e^{-\frac12\vf^--\frac32\vf^+} k^{--}S^{+-}S^{++} \cr
v_h=1:&\quad  c\,e^{-\frac12\vf^--\frac32\vf^+} k^{--}S^{+-}S^{++} \cr
v_h=2:&\quad  c\pa c\,e^{-\frac12\vf^--\frac32\vf^+} k^{--}S^{+-}S^{++} \cr
v_h=2':&\quad \tilde cc\,e^{-\frac12\vf^--\frac32\vf^+} k^{--}S^{+-}S^{++} \cr
v_h=3:&\quad  \tilde cc\pa c\,e^{-\frac12\vf^--\frac32\vf^+}
k^{--}S^{+-}S^{++} \cr
}\eqno(A.6b)$$

$$\eqalign{ &(\p_+,\p_-)=(-\fracm32,-\fracm12):
\qquad\qquad\qquad\qquad\qquad\qquad\qquad\qquad\cr
v_h=0:&\quad  \int\,e^{-\frac32\vf^--\frac12\vf^+} k^{+-}S^{-+}S^{--} \cr
v_h=1:&\quad  c\,e^{-\frac32\vf^--\frac12\vf^+} k^{+-}S^{-+}S^{--} \cr
v_h=2:&\quad  c\pa c\,e^{-\frac32\vf^--\frac12\vf^+} k^{+-}S^{-+}S^{--} \cr
v_h=2':&\quad \tilde cc\,e^{-\frac32\vf^--\frac12\vf^+} k^{+-}S^{-+}S^{--} \cr
v_h=3:&\quad  \tilde cc\pa c\,e^{-\frac32\vf^--\frac12\vf^+}
k^{+-}S^{-+}S^{--} \cr
}\eqno(A.7b)$$

$$\eqalign{ &(\p_+,\p_-)=(-\fracm12,-\fracm12):
\qquad\qquad\qquad\qquad\qquad\qquad\qquad\qquad\cr
v_h=0:&\quad  \int\,e^{-\frac12\vf^--\frac12\vf^+}
[k^{\pm-}S^{-+}S^{++}\pm ik^{\pm+}S^{+-}S^{--}] \cr
v_h=1:&\quad  c\,e^{-\frac12\vf^--\frac12\vf^+}
[k^{\pm-}S^{-+}S^{++}\pm ik^{\pm+}S^{+-}S^{--}] \cr
v_h=2:&\quad  c\pa c\,e^{-\frac12\vf^--\frac12\vf^+}
[k^{\pm-}S^{-+}S^{++}\pm ik^{\pm+}S^{+-}S^{--}]\cr
&\qquad +c\h^\mp\,e^{\pm\frac12\vf^-\mp\frac12\vf^+}
S^{\pm\mp}S^{\pm\pm}/k^{\mp-} \cr
v_h=2':&\quad \tilde cc\,e^{-\frac12\vf^--\frac12\vf^+}
[k^{\pm-}S^{-+}S^{++}\pm ik^{\pm+}S^{+-}S^{--}] \cr
v_h=3:&\quad  \tilde cc\pa c\,e^{-\frac12\vf^--\frac12\vf^+}
[k^{\pm-}S^{-+}S^{++}\pm ik^{\pm+}S^{+-}S^{--}]\cr
&\qquad +\tilde cc\h^\mp\,e^{\pm\frac12\vf^-\mp\frac12\vf^+}
S^{\pm\mp}S^{\pm\pm}/k^{\mp-} \cr
}\eqno(A.8b)$$

\end{document}